\documentclass[twocolumn,superscriptaddress,aps,prb,10pt]{revtex4-2}
\usepackage{amsfonts}
\usepackage{amssymb} 
\usepackage{subfigure}
\usepackage{picinpar}
\usepackage[marginal]{footmisc}
\usepackage{braket}
\usepackage{bm}
\usepackage{setspace}
\usepackage{amsmath,makecell}
\usepackage{graphicx}
\usepackage[colorlinks,linkcolor=blue,citecolor=blue,urlcolor=blue]{hyperref}
\usepackage{adjustbox} 
\usepackage{ragged2e}
\usepackage{placeins} 
\usepackage{booktabs}
\usepackage{tabularx}
\usepackage{lipsum} 
\usepackage{afterpage}
\usepackage{threeparttable}
\makeatletter

\newcommand{\Rmnum}[1]{\expandafter\@slowromancap\romannumeral #1@}

\makeatother

\begin{document}
\title{Tuning Topological States by Dissipation}
\author{ Xue-Ping Ren\quad}
\thanks{Who has same contribution to this work}
\affiliation{Center for Advanced Quantum Studies, School of Physics and Astronomy, Beijing Normal University, Beijing 100875, China}
\affiliation{Key Laboratory of Multiscale Spin Physics (Ministry of Education), Beijing Normal University, Beijing 100875, China}
\author{Yue Hu }
\thanks{Who has same contribution to this work}
\affiliation{Center for Advanced Quantum Studies, School of Physics and Astronomy, Beijing Normal University, Beijing 100875, China}
\affiliation{Key Laboratory of Multiscale Spin Physics (Ministry of Education), Beijing Normal University, Beijing 100875, China}
\author{ Long-Ye Lu}
\affiliation{Center for Advanced Quantum Studies, School of Physics and Astronomy, Beijing Normal University, Beijing 100875, China}
\affiliation{Key Laboratory of Multiscale Spin Physics (Ministry of Education), Beijing Normal University, Beijing 100875, China}
\author{ Xin-Ran Ma}
\affiliation{Center for Advanced Quantum Studies, School of Physics and Astronomy, Beijing Normal University, Beijing 100875, China}
\affiliation{Key Laboratory of Multiscale Spin Physics (Ministry of Education), Beijing Normal University, Beijing 100875, China}
\author{Ji-Yao Fan}
\affiliation{Center for Advanced Quantum Studies, School of Physics and Astronomy, Beijing Normal University, Beijing 100875, China}
\affiliation{Key Laboratory of Multiscale Spin Physics (Ministry of Education), Beijing Normal University, Beijing 100875, China}
\author{Cui-Xian Guo}
\affiliation{Beijing Key Laboratory of Optical Detection Technology for Oil and Gas, China University of Petroleum-Beijing, Beijing 102249, China}
\affiliation{Basic Research Center for Energy Interdisciplinary, College of Science, China University of Petroleum-Beijing, Beijing 102249,China}
 \author{Su-Peng Kou}
\email[Email: ]{spkou@bnu.edu.cn}
\affiliation{Center for Advanced Quantum Studies, School of Physics and Astronomy, Beijing Normal University, Beijing 100875, China}
\affiliation{Key Laboratory of Multiscale Spin Physics (Ministry of Education), Beijing Normal University, Beijing 100875, China}
\date{\today}
\begin{abstract}
The bulk-boundary correspondence plays a crucial role in topological quantum systems, however, this principle is broken in non-Hermitian systems. The breakdown of the bulk-boundary correspondence indicates that the global phase diagrams under open boundary conditions are significantly different from those under periodic boundary conditions. In this paper, we investigate how the bulk-boundary correspondence breaks down by gradually tearing the system. We find that by tuning the strength of gain and loss domain wall, in the thermodynamic limit, the global phase diagrams of the topological system become the hybrids of those under periodic and open boundary conditions. Moreover, during the breakdown of the bulk-boundary correspondence, several phase transitions occur. This situation is quite different from earlier work, where the breakdown of the bulk-boundary correspondence in the thermodynamic limit occurred suddenly due to infinitesimal boundary hopping amplitudes. To support our conclusions, we provide both analytical and numerical calculations. These results help researchers better understand non-Hermitian topological systems.
\end{abstract}
\maketitle
\section{Introduction}\label{sec:intro}  
Non-Hermitian systems exhibit a plethora of intriguing phenomena different from Hermitian systems, including the non-Hermitian skin effect (NHSE) \cite{PhysRevLett.121.086803,PhysRevA.104.022215,annurev:/content/journals/10.1146/annurev-conmatphys-040521-033133,doi:10.34133/2021/5608038,PhysRevLett.123.170401,0Impurity}, broken bulk-boundary correspondence (BBC) \cite{PhysRevB.103.075126,2020Non,2020Bulk,PhysRevB.99.081103,PhysRevB.99.081302,PhysRevLett.121.026808}, complex eigenspectra, complex energy gaps, and exceptional points \cite{RevModPhys.93.015005,PhysRevLett.123.066405,2019Parity,doi:10.1126/science.aar7709,PhysRevB.95.184306,PhysRevA.100.062131,PhysRevLett.118.045701}, and so on.  To better describe and explain the behavior of non-Hermitian systems, researchers have proposed many new methods and concepts, such as GBZ \cite{PhysRevLett.121.086803,PhysRevLett.121.136802}, parity-time (PT) symmetry \cite{PhysRevLett.80.5243,10.1063/1.532860,PhysRevLett.89.270401,PhysRevLett.100.103904,El-Ganainy:07,PhysRevLett.100.030402,PhysRevLett.103.093902,R2010Observation}, and non-Hermitian topological states \cite{2018Topological,PhysRevB.99.201103,Ghatak_2019,PhysRevLett.124.056802}.
Non-Hermiticity is typically achieved through the introduction of nonreciprocal hopping terms \cite{PhysRevB.92.094204,PhysRevB.95.014201,PhysRevLett.123.170401,doi:10.1073/pnas.2010580117,2020Generalized} or non-Hermitian gain and loss terms \cite{PhysRevB.97.075128,PhysRevA.89.062102,PhysRevApplied.13.014047,PhysRevLett.123.165701,PhysRevB.97.045106}.
There has been a growing interest in topological
properties of non-Hermitian Hamiltonians \cite{PhysRevB.84.205128,PhysRevA.87.012118,PhysRevLett.116.133903,PhysRevLett.118.040401} applicable
to a wide range of systems such as systems with open boundaries \cite{Rotter_2009,PhysRevLett.104.153601,RevModPhys.87.61} and systems with gain and/or loss \cite{PhysRevLett.121.136802,PhysRevLett.80.5243,PhysRevLett.101.080402}.
For instance, with suitable tuning of the gain and loss quantity, there can be topological phase transition of the photonic bands from topological trivial phase to non-trivial phase \cite{ZHOU2020125653,PhysRevA.104.033501,PhysRevA.105.053510,Jiang:22,PhysRevA.105.023531,Wang:23}.
 Ref. \cite{PhysRevLett.129.053903} propose a universal method to
construct effective Hamiltonians in gain and loss domain walls system. They can give faithful descriptions of the eigenenergies and field distributions of
domain induced states. The non-Hermitian tearing theory is proposed for gain and loss systems in Ref. \cite{2024tearing}. They pointed out that the gain and loss domain wall have a tearing effect, causing the separation of bulk states and the decoupling of boundary states, and proposed an effective model to explain this phenomenon.

The breakdown of BBC originates from the sensitivity of non-Hermitian systems to boundary, under different boundary conditions \cite{2024tearing,2020Non,PhysRevB.100.035102,PhysRevLett.127.116801,Xiong2018Why}, the system can exhibit various novel physical phenomena.  The conventional approach of eliminating boundary hopping amplitudes \cite{PhysRevLett.127.116801} to achieve the open boundary conditions (OBC) leads to an abrupt breakdown of BBC in the thermodynamic limit --- where even infinitesimal boundary hopping amplitudes are sufficient to trigger this phenomenon. Appropriate gain and loss domain wall could potentially substitute for the OBC \cite{2020Non}. However, in gain and loss domain wall system, during the process of tuning the  dissipation strength, might the system exhibit some intriguing phenomena distinct from those observed in conventional approaches of eliminating boundary hopping amplitudes to OBC?  No research has been conducted on this topic. What are the characteristic transformations of the phase diagram throughout this process?  Furthermore, what is the physical mechanism underlying the breakdown of BBC in gain and loss domain wall?  

In this paper, we investigated how dissipation affects boundary condition and tunes topological states. Considering a one-dimensional (1D) periodic Su-Schrieffer-Heeger (SSH) chain \cite{PhysRevB.22.2099} with different nonreciprocal intracell couplings as is shown in Fig. \ref{fig:1}, adding gain and loss into left and right sublattice, respectively, forming a domain wall. As the dissipation increases, system gradually evolves from the periodic boundary conditions (PBC) to the OBC, then one system with PBC is torn into two systems with OBC ultimately, which can be convincingly demonstrated by the excellent agreement between numerical and analytical results. In this process, the phase diagrams exhibit the hybrids of those under PBC and OBC, whose topological regions are both dissipation-tunable and analytically solvable, reflecting the continuity of the BBC breaks down. Notably, this mechanism fundamentally differs from the conventional approach of eliminating boundary hopping amplitudes to achieve OBC, under which the breakdown of BBC occurs abruptly due to  due to infinitesimal boundary hopping amplitudes in the thermodynamic limit.  In addition, during the breakdown of BBC, the richer phases and phase transition emerge, which are defined by their energy spectrum structures.  Interestingly, each phase transition corresponds to distinctive behavior in the GBZ, presenting the rich configuration of the GBZ. Our work  help researchers better understand non-Hermitian topological systems.

The article is organized as follows: In Sect. \ref{sec:method}, we introduce the theoretical model. In Sect. \ref{sec:Evolution},  we demonstrate how the phase diagram evolves during the process of opening boundary in different ways. In Sect. \ref{sec:Continuous}, we characterize the various phase signature and phase transition during the breakdown of BBC, and determine analytically the boundary lines of topological region under  dissipation-tuned.  In Sect. \ref{sec:conclu}, we draw the conclusion.

\section{Model}\label{sec:method}
\begin{figure}[htbp]
\begin{minipage}{0.5\textwidth}
\vspace*{-0.5cm} 
\centering
\includegraphics[scale=0.3]{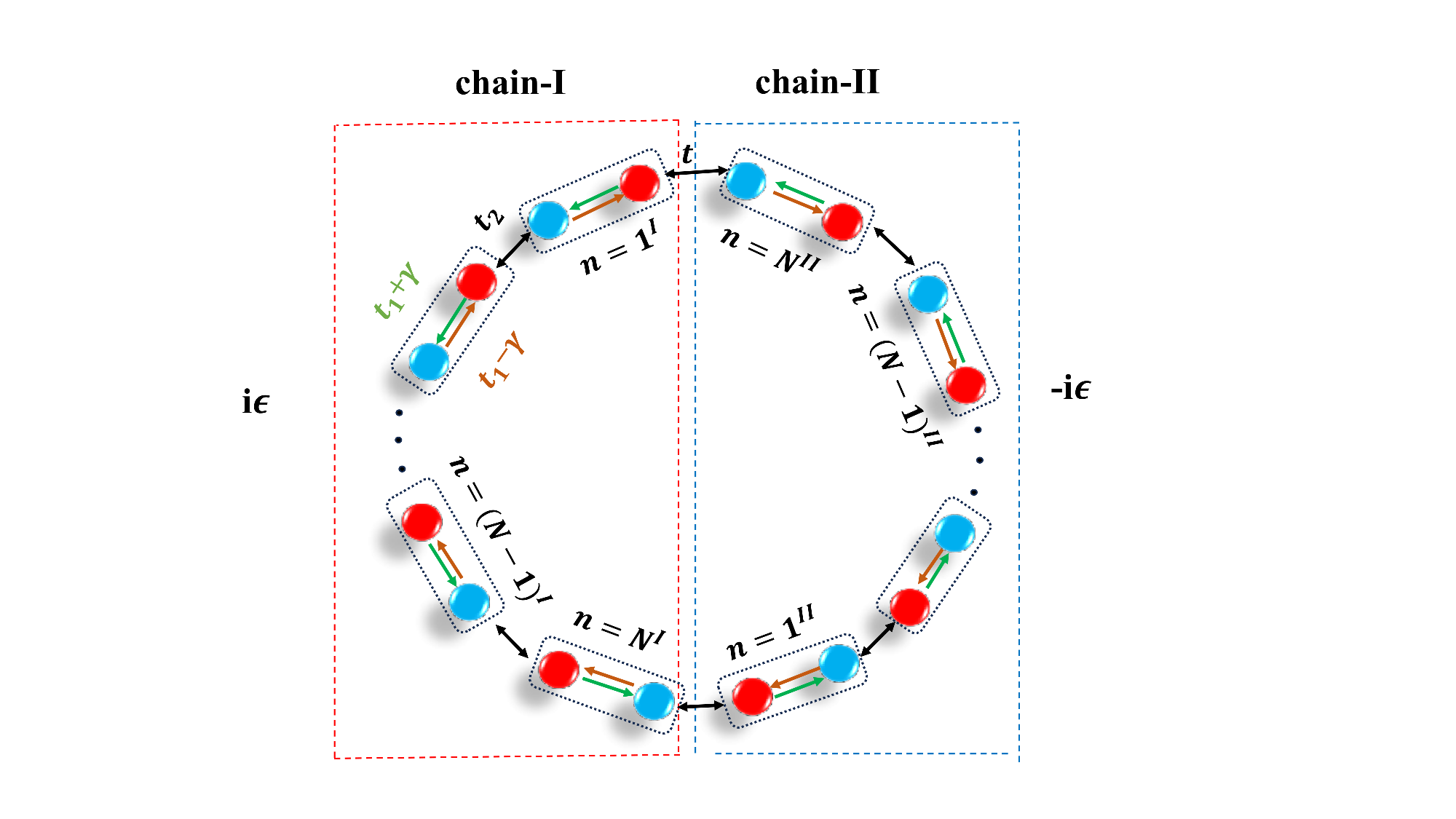}
\caption{Schematic depiction of a 1D periodic SSH chain with  directional nonreciprocal intracell couplings, adding on-site gain '$\rm i\epsilon$' and loss '$- \rm i\epsilon$'  to all sublattice of chains I and II, respectively. Chains I and II indicate cells that from $\rm 1^{\rm I/ \rm II }$ to $\rm N^{\rm I/ \rm II }$, respectively. Here, $ \rm N^{\rm I} = N^{\rm II} = \rm N$.  The  nonreciprocal intracell coupling is  $t_1 \pm \gamma$, and intercell hopping is $t_2$, and $t$ indicate boundary hopping amplitudes.}
\label{fig:1}
\end{minipage}
\end{figure}

We consider a one-dimensional (1D)  periodic  SSH chain with different nonreciprocal intracell couplings as is shown in Fig. \ref{fig:1}. The on-site gain '$\rm i\epsilon$' and loss '$-\rm i\epsilon$' are introduced to all sublattice  of chains I and II, respectively, forming a gain and loss domain wall described by $V ={ \rm i } \epsilon \sum_{n=1^{\rm I}}^{N^{\rm I}} ( c^{\dag}_{n,a} c_{n,a}+c^{\dag}_{n,b} c_{n,b}  )
- {\rm i \epsilon} \sum_{n=1^{\rm II}}^{N^{\rm II}} (c^{\dag}_{n,a} c_{n,a}+ c^{\dag}_{n,b} c_{n,b})$. In our model,  an equal number of unit cells in both chains is considered, i.e., $ \rm N^{\rm I} = N^{\rm II}$ = N. The strength of intercell coupling is represented by $t_2$, while the directional nonreciprocal intracell coupling is represented by  $t_1\pm \gamma$ with $\gamma$ nonreciprocity strength, where $\pm$ denotes  left and right directions, respectively. The boundary  hopping amplitudes are denoted by $t$.

The whole system can be viewed as a coupling of chains I and II, therefore, the Hamiltonian has the following
matrix form in real space:
\begin{equation}\label{eq:1}
 H= 
   \begin{pmatrix}
      H^{\rm I} & t_c \\
      t^{\top}_{c} & H^{\rm II} 
    \end{pmatrix},
\end{equation}
where, $t_c$ is a matrix of size $2\rm N \times 2 \rm N $ and all the elements are $0$ except the upper right and lower left corners.  $H^{\rm  I}$ and $H^{\rm II}$ denote the Hamiltonians for chains I and II, respectively. The Bloch Hamiltonian for the uncoupled chains with translation invariance read:
\begin{equation}\label{eq:2}
    H^{\rm \rm I/  \rm II} (k) = d_x \sigma_x + (d_y + {\rm i} \gamma) \sigma_y \pm {\rm i \epsilon} \sigma_0,
\end{equation}
where, $d_x=t_1+t_2 {\rm cos(k)}, d_y=t_2 {\rm sin(k)}, \sigma_{x,y}$ is Pauli matrix, $\sigma_0$ is the identity matrix, and $\pm$ indicates chains I and II, respectively. The real-space eigenvalue
equation for the coupled chain Hamiltonian $H$ takes the form $H \Psi = E \Psi$, where $\Psi=(\Psi^{\rm I}_{1 A}, \Psi^{\rm I}_{1 B}, \cdots, \Psi^{\rm I}_{N A},\Psi^{\rm I}_{N B},\Psi^{\rm II}_{1 A}, \Psi^{\rm II}_{1 B}, \cdots, \Psi^{\rm II}_{N A},\Psi^{\rm II}_{N B})^{\top}$ denotes the associated eigenfunction and $E$ represents the corresponding energy eigenvalue.

The model possess translation invariance when $\epsilon=0$, then the eigenenergy is:
\begin{equation}\label{eq:3}
E(k)= \pm \sqrt{(t_1 + t_2 {\rm cos(k)})^2 + (t_2  {\rm sin(k)} + {\rm i} \gamma)^2}.
\end{equation}
\section{Evolution of phase diagrams}\label{sec:Evolution}
\begin{figure*}[ht]
 \vspace*{-2cm}
 \hspace*{-4cm}
    \includegraphics[scale=4.5]{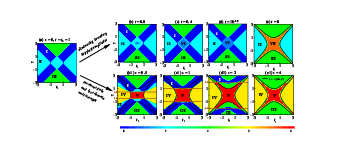}
 \vspace*{-2cm}
\caption{Phase diagram where the boundary is opened in different ways for 1D peridorc SSH model. (a)-(e) The phase diagram change with gradual attenuation of boundary hopping amplitudes  $t$. (a)-(e1) The phase diagram change with the increase of the dissipation domain wall $\epsilon$. $ \rm{\rm I}$ (dark blue) gapless,  $ \rm{\rm II}$ (light blue) and $ \rm{\rm VI}$ (medium blue)  real gap opens,  $ \rm{\Rmnum{3}}$ (green) imaginary gap opens, $ \rm{\Rmnum{4}}$  (yellow) the real and imaginary gap open simultaneously, $\rm{\Rmnum{5}}$  (red) the real and imaginary gap open simultaneously and there are topological boundary states, $\rm{\Rmnum{7}}$  (orange) real gap opens, accompanied by the emergence of topological boundary states. The solid black lines are the topological phase boundary \cite{PhysRevLett.121.086803}. The dashed brown lines is $\epsilon=\pm \gamma$. Common parameters: $ t_2=1$.}
\label{fig:2}
\end{figure*}

In this subsection, we explore how the system's phase diagram evolves as the boundary is opened in various ways. Traditionally, by gradually eliminating boundary hopping amplitudes $t$ to reach OBC \cite{PhysRevLett.127.116801,Xiong2018Why}. As this boundary hopping amplitudes decays, the phase diagram undergoes a series of transformations, which we categorize based on the structure of the energy spectrum. Fig. \ref{fig:2}(a)-(e) illustrate this process. Here, region I (dark blue) denotes a gapless phase. Region II (light blue) and region VI (medium blue) both signify the opening of a real gap, however,  the key the difference is that region VI hosts in-gap bound states, as described  in Ref. \cite{PhysRevLett.127.116801}.  As the boundary hopping amplitudes decays for $ t \neq 0$, these bound states gradually emerge from (b) to (d), and their localization becomes increasingly pronounced, and the NHSE vanishes for $t\neq0$  in the thermodynamic limit. Region III (green) indicates the opening of an imaginary gap. Region VII (orange) represents a real gap with topological edge states, which corresponds to the topological phase under OBC. Regions II, III, VI and VII all feature an energy spectrum split into two parts in the complex plane. Fig. \ref{fig:2}(b)-(d) depict  the boundary conditions that lie between PBC and OBC, which we refer to as 'incomplete PBC' (iPBC). The phase diagram under iPBC approach to that under PBC, exhibiting the same properties. It is worth noting that in the thermodynamic limit, under the infinitesimal boundary hopping amplitudes $t=10^{-20}$, the phase diagram undergoes a sudden change, accompanied by the breakdown of the BBC, as shown in (d) and (e). 

Unlike the traditional ways of opening the system's boundary, opening the boundary by tuning the gain and loss domain wall strength can give rise to more interesting physical phenomena, which is also the focus of our discussion in this paper. Fig. \ref{fig:2}(a)-(e1) show the process in which the system's boundary is opened as the dissipation gradually increases,  presenting a series of much richer phase diagram, which feature the more abundant energy spectrum structure compared to Fig. \ref{fig:2}(a)-(e). Interestingly, the phase diagrams under iPBC exhibit the states of hybrid, which are the hybridization of the phase diagrams under PBC and OBC with the tuning of dissipation, as shown in Fig. \ref{fig:2}(b1)-(d1), where regions I-III are identical to those depicted in Fig. \ref{fig:2}(a)-(e), sharing the same energy spectrum structure. Region IV (yellow) indicates that the real and imaginary gaps open simultaneously, dividing the complex plane into four parts. The distinction between region $\rm \Rmnum {5}$ and region $\rm \Rmnum {4}$ lies in the appearance of topological edge states within region $\rm \Rmnum {5}$, which corresponds to the topological phase.

As dissipation strengthen, region $\rm \Rmnum {5}$ progressively expands along the OBC topological boundary (solid black line) \cite{PhysRevLett.121.086803} until it completely saturates the topological phase space. This expansion is a continuous process, and under iPBC,  the boundary line of the topological region need to satisfy two conditions simultaneously: $t_1=\pm \sqrt{t_2^2+\gamma^2} $ and $ \epsilon=\pm\gamma$ (the brown dashed line), which will be further explained in In Sect. \ref{subsec:IVC}. Simultaneously, region $\rm \Rmnum {4}$ gradually covers region $\rm \Rmnum {2}$ and part of region $\rm \Rmnum {1}$ along the boundary line of the topological region, eventually saturates the left and right parts of the non-topological region.  Region $\rm \Rmnum {1}$ gradually shrinks and moves up and down until it completely disappears. Region $\rm \Rmnum {3}$ gradually covers part of region $\rm \Rmnum {1}$ and expands along the topological phase boundary until it saturates the upper and lower parts of the non-topological region. This phase diagram hybridization process indicates that the originally gapless and single-real-gap energy spectrum are torn under dissipation tuning, which is because the tearing effect of the dissipation domain wall tends to shift the energy spectrum up and down along the imaginary axis, effectively dividing the complex plane into upper and lower parts \cite{2024tearing}. Interestingly, this continuous change process is independent of the system size. This means that in the thermodynamic limit, the phase diagram undergoes a series of continuous changes, accompanied by the continuous breakdown of the BBC, which is different from the scenario of slowly  eliminating boundary hopping amplitudes. To more intuitively illustrate the differences between these two methods, we present a table as shown in Tab.\ref{tab:1}.

\begin{table}[t] 
\begin{threeparttable}
\centering
  \caption{Topological region under different boundary condition tunings}
  \label{tab:1}
\begin{ruledtabular}
\begin{tabular}{ccc}
\addlinespace[12pt]
&\makecell {Tuning the boundary\\ hopping $'t'$ amplitudes}      &\makecell{Tuning Gain and loss \\ domain wall $'\epsilon'$ strength} \\   
\addlinespace[12pt]
\colrule
\addlinespace[12pt]
 &  \makecell{ $  \begin{aligned} &t=0: t_1=\pm \sqrt{t_2^2+\gamma^2} \\ & t=0^{+}: t_1=|t_2 \pm  \gamma| \end{aligned} $ } 
 \,
 &  \makecell{$ \begin{aligned}  & \epsilon > |\gamma|:  t_1=\pm \sqrt{t_2^2+\gamma^2}  \\  &   \epsilon < |\gamma|:  t_1 \approx |t_2 \pm  \gamma|    \end{aligned}$  }     \\
 \addlinespace[12pt]
\end{tabular}
\end{ruledtabular}
\end{threeparttable}
\end{table}

\begin{figure}[t]
  \includegraphics[scale=1.5]{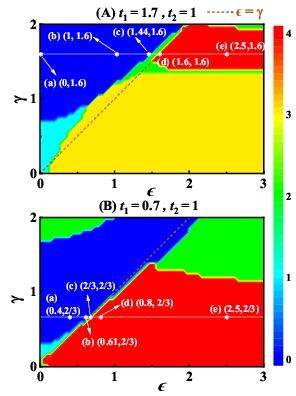}
  \caption{The phase diagram on the parameter space of $\epsilon$ and $\gamma$ for the SSH model with $t_1=1.7$ and $0.7$, respectively. The various regions correspond to Fig. \ref{fig:2}(a)-(e1). The dashed brown lines represent $\epsilon=\gamma$. Common parameters: $\rm 2N=60$, $t_2=1$.}
  \label{fig:3}
\end{figure}

To systematically investigate how dissipation tunes topological states, we present phase diagrams in the parameter space of $ \gamma \, \& \, \epsilon$ with $t_1=1.7$ and 0.7 in Fig. \ref{fig:3}. The various colored regions correspond to Fig. \ref{fig:2}(a)-(e1). The appearance of topological states (red region) remains our central concern. The white lines connecting points (a)-(e) in Fig. \ref{fig:3} illustrate that the energy spectrum structure undergoes a series of transformations. For large $|t_1|$ and $\gamma$  in Fig. \ref{fig:3}(A), as dissipation increases, the energy spectrum evolves through several phases: it starts as gapless phase (region I), then develops an imaginary gap (region III), and finally both real and imaginary gaps open simultaneously, accompanied by the emergence of topological edge states (region V), as indicated by the white line in Fig. \ref{fig:3}(A). For small $|t_1|$  and $\gamma$  in Fig. \ref{fig:3}(B), the energy spectrum undergoes a different sequence transformation as dissipation increases: it begins as gapless phase (region I), then forms a real gap (region II), and eventually both real and imaginary gaps open simultaneously, with the appearance of topological edge states (region V), as shown by the white line in Fig. \ref{fig:3}(B). These dual transformations share the universal feature that the concomitant real and imaginary gap openings are invariably tied to topological boundary modes, whose emergence excellent agreement with the $\epsilon=\gamma$ (the brown dotted line) condition. A systematic characterization of the phases and their transformation pathways is presented in subsequent sections. 

\section{Continuous process of BBC breakdown}\label{sec:Continuous}
\subsection{Theoretical method}\label{subsec:IVA}
To investigate the tuning role of the gain and loss domain wall, we employed a combination of numerical and analytical methods and obtained analytical solutions by calculating the GBZ equation of the model. First, by solving the Schrödinger equation in real space $H \Psi = E \Psi$ (see Appendix \ref{app:A} for detailed derivation), we derived the characteristic equations for the two chains as follows:
\begin{equation} \label{eq:4}
(E^{\rm I / \rm II})^2=t_1^2-\gamma^2+t_2^2+t_2(\beta^{\rm I / \rm II})^{-1}(t_1-\gamma)+t_2 \beta^{\rm I  / \rm II} (t_1+\gamma),
\end{equation} 
where, $E^{\rm I / \rm II}= E \mp i\epsilon $ and $\beta^{\rm I / \rm II}$ denote the eigenvalues and GBZ solutions of chains I and II, respectively.

Then, to satisfy the boundary conditions, $\beta^{\rm I / \rm II}$ in the characteristic equation is restricted  as follows (see Appendix \ref{app:B} for detailed derivation):
\onecolumngrid
\begin{equation}\label{eq:5}
 \begin{aligned}
 & [(\beta^{I}_1)^N (\beta^{\rm II}_1)^N-1][(\beta^{I}_2)^N (\beta^{\rm II}_2)^N-1](\eta^{I}_1 \beta^{\rm I}_1 -\eta^{\rm II}_2 \beta^{\rm II}_2)(\eta^{\rm I}_2 \beta^{\rm I}_2 -\eta^{\rm II}_1 \beta^{\rm II}_1) \\
  =&[(\beta^{\rm I}_2)^N (\beta^{\rm II}_1)^N-1][(\beta^{\rm I}_1)^N (\beta^{\rm II}_2)^N-1]
  (\eta^{\rm I}_1 \beta^{\rm I}_1 -\eta^{\rm II}_1 \beta^{\rm II}_1)(\eta^{\rm I}_2 \beta^{\rm I}_2 -\eta^{\rm II}_2 \beta^{\rm II}_2) .
 \end{aligned}
\end{equation}
\twocolumngrid
Finally, in the thermodynamic limit $(N \rightarrow \infty)$, the GBZ equation for this model can be obtained by reducing the restriction Eq.(\ref{eq:5}) (see Appendix \ref{app:C} for detailed derivation):
\begin{equation}\label{eq:6}
0=
  \begin{cases}
    \lvert \beta ^{\rm I}_1 \beta ^{\rm II}_1 \rvert -1 ,  &\textsl{g}_1 \geq 1 \wedge  \textsl{g}_2 \geq 1 \wedge  \textsl{g}_3 \geq 1 \\
    \lvert \beta ^{\rm I}_1 \rvert -\lvert \beta ^{\rm I}_2 \rvert, & \textsl{g}_1 \geq \textsl{g}_3 \geq 1  \geq  \textsl{g}_2 \geq \textsl{g}_4  \\
    \lvert \beta ^{\rm II}_1 \rvert - \lvert \beta ^{\rm II}_2 \rvert, & \textsl{g}_1 \geq \textsl{g}_2 \geq 1 \geq \textsl{g}_3 \geq \textsl{g}_4   \\
    \lvert \beta ^{\rm I}_2 \beta ^{\rm II}_2 \rvert -1,   &\textsl{g}_2 \leq 1 \wedge \textsl{g}_3 \leq 1  \wedge  \textsl{g}_4 \leq 1 ,\\
  \end{cases}
\end{equation}
then, detailed calculations of GBZ solutions $\beta ^{\rm I/II}_{1/2}$ under different conditions are given in Appendix \ref{app:D}.
\subsection{Characteristics of various phases}\label{subsec:IVB}

\begin{figure*}[htbp]
    \centering
        \begin{minipage}[b]{0.18\linewidth}
        \includegraphics[scale=0.19]{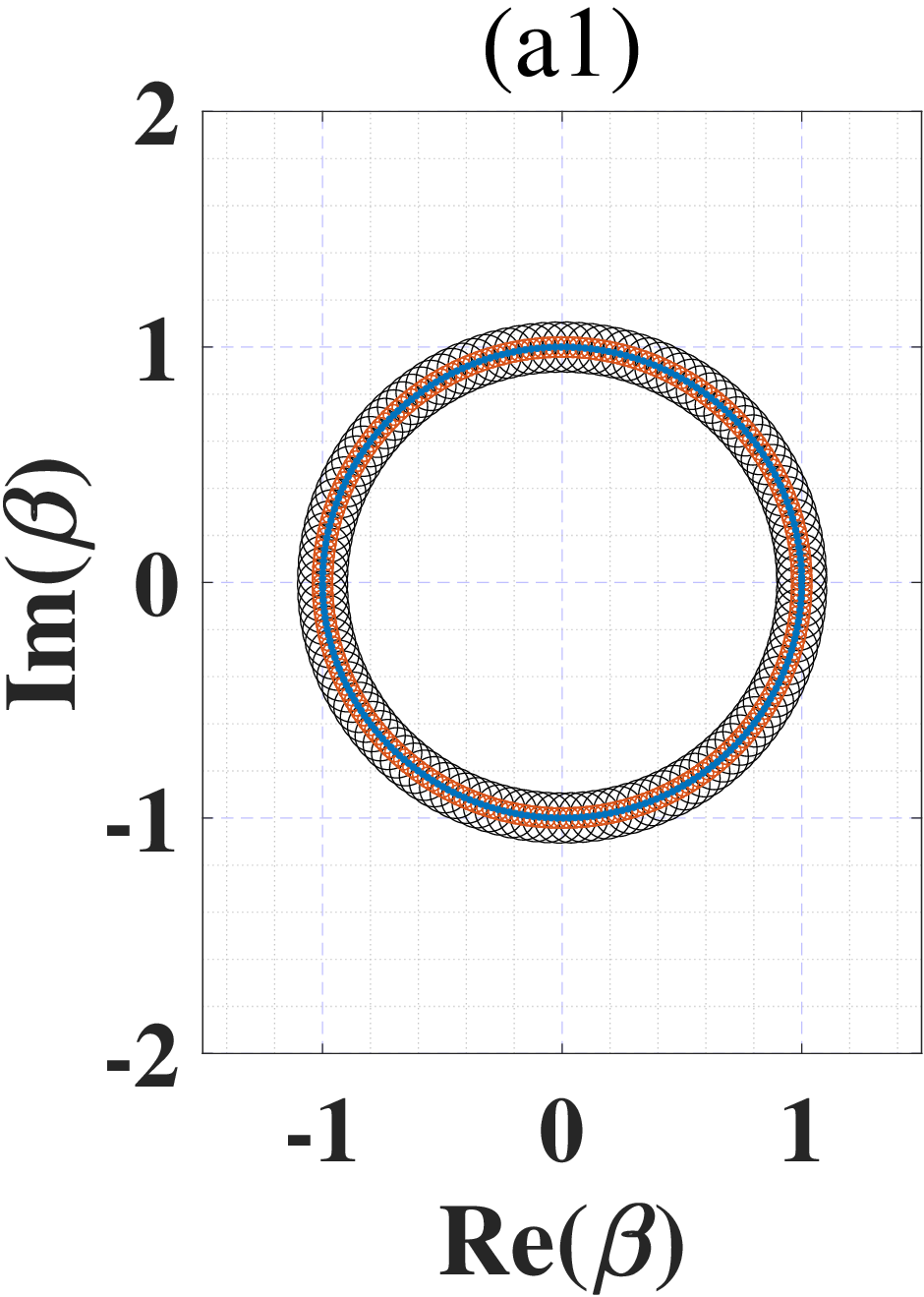}
    \end{minipage}
     \hspace{0.2cm}
    \begin{minipage}[b]{0.18\linewidth}
       \includegraphics[scale=0.23]{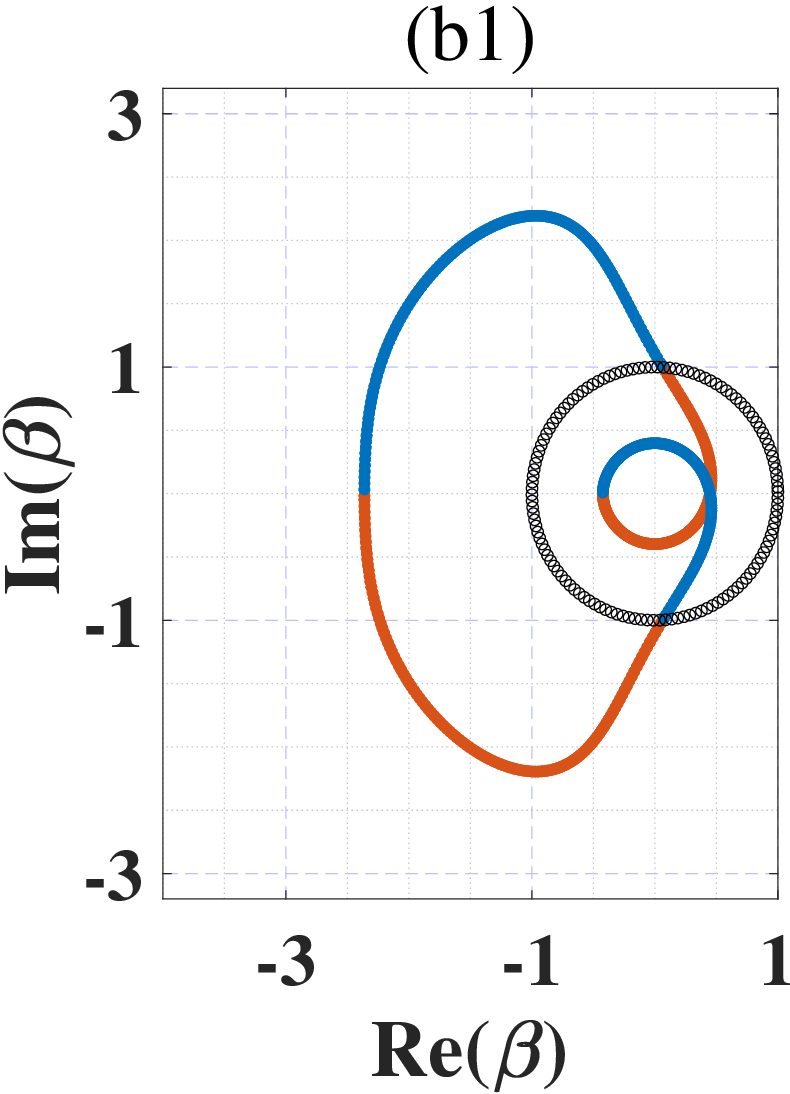}
    \end{minipage}
        \hspace{0.2cm}
    \begin{minipage}[b]{0.18\linewidth}
       \includegraphics[scale=0.23]{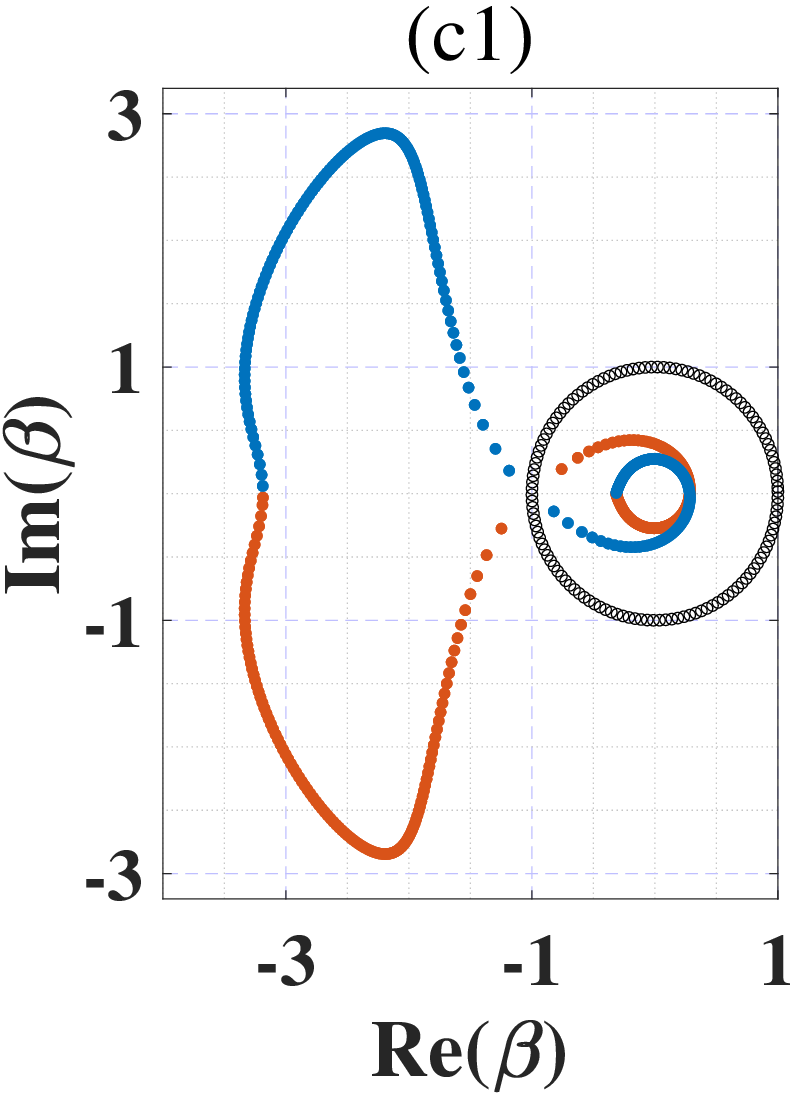}
    \end{minipage}
          \hspace{0.2cm}
    \begin{minipage}[b]{0.18\linewidth}
       \includegraphics[scale=0.23]{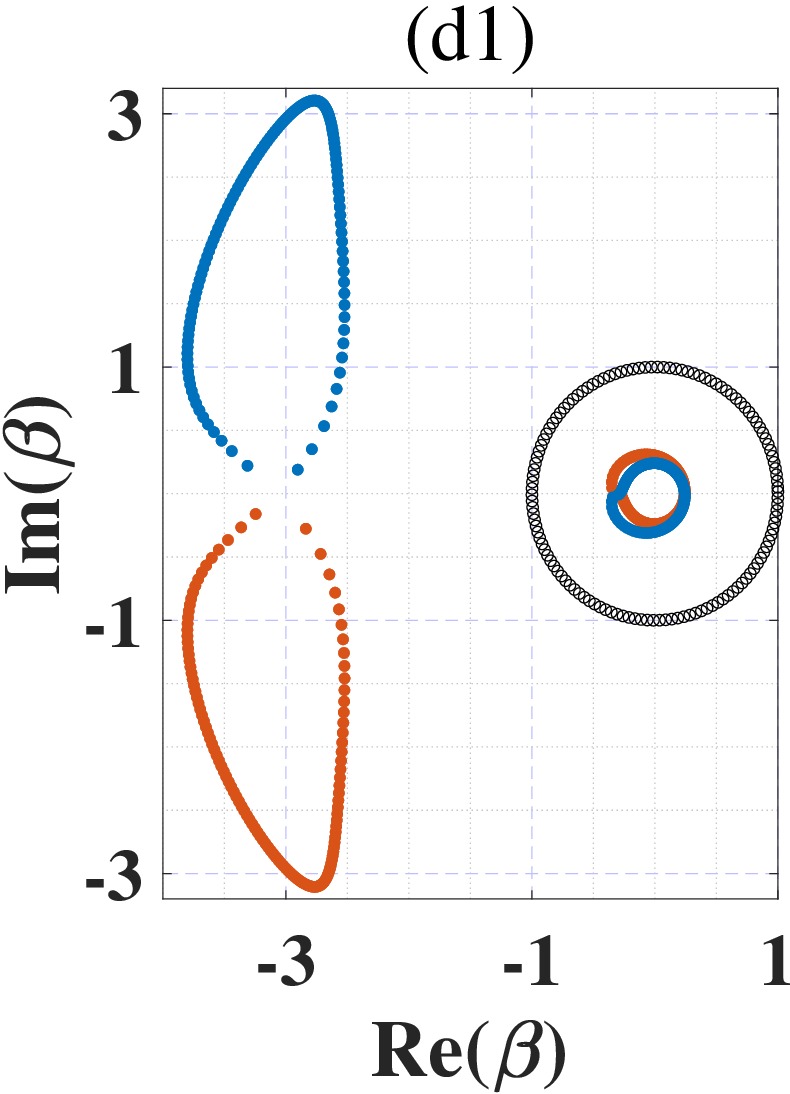}
    \end{minipage}
       \hspace{0.2cm}
    \begin{minipage}[b]{0.18\linewidth}
       \includegraphics[scale=0.19]{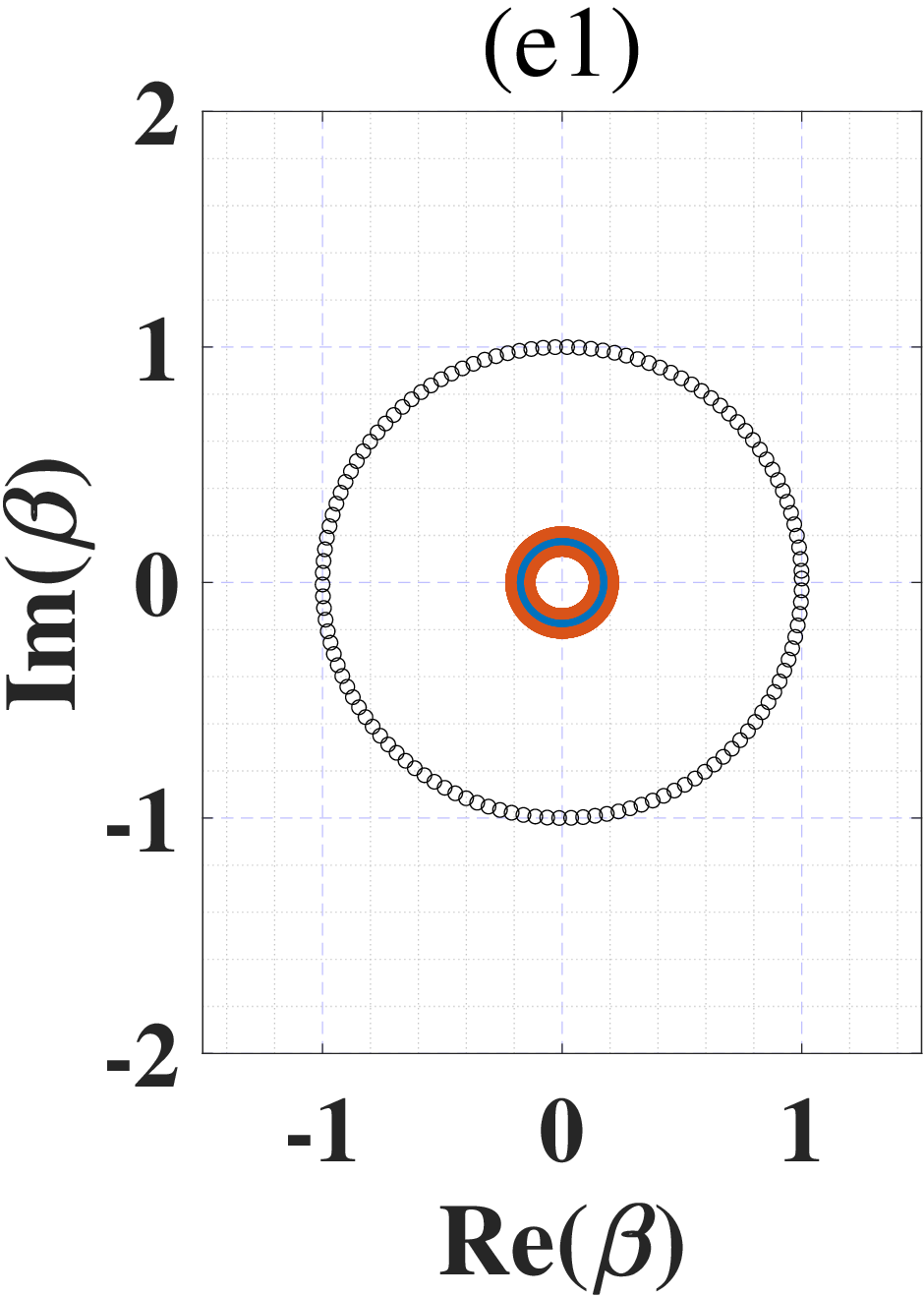}
    \end{minipage}
    \vspace{0.3cm}
    \begin{minipage}[b]{0.18\linewidth}
       \includegraphics[scale=0.235]{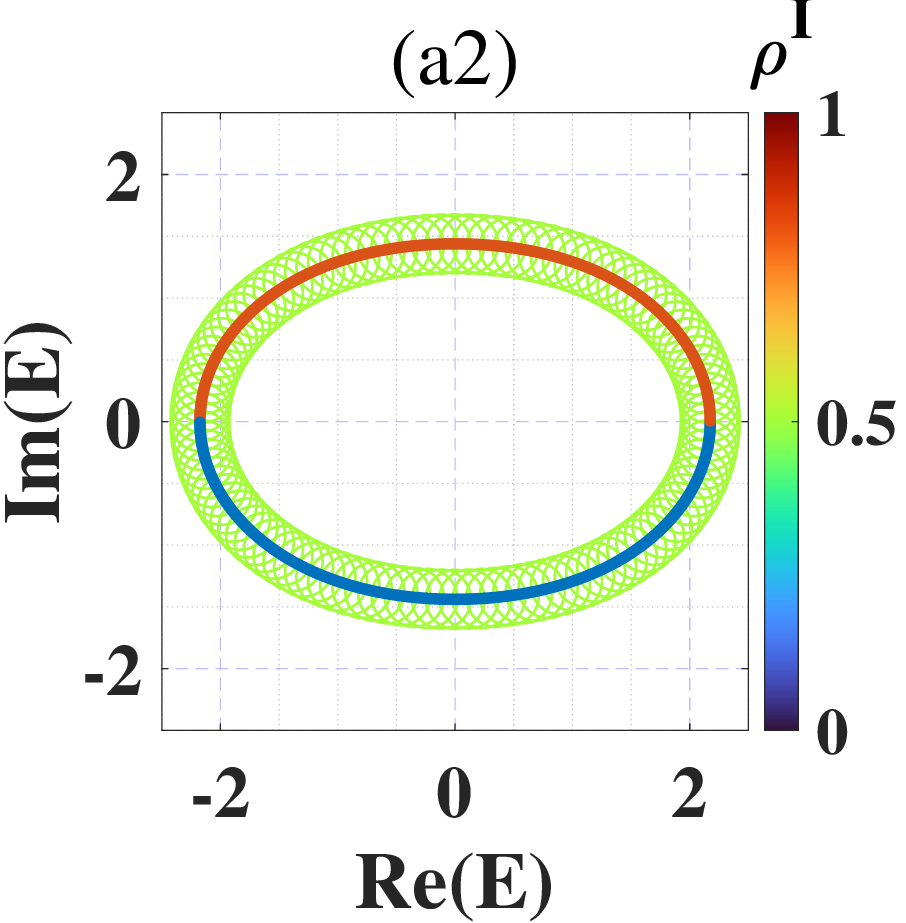}
    \end{minipage}
       \hspace{0.15cm}
    \begin{minipage}[b]{0.18\linewidth}
        \includegraphics[scale=0.235]{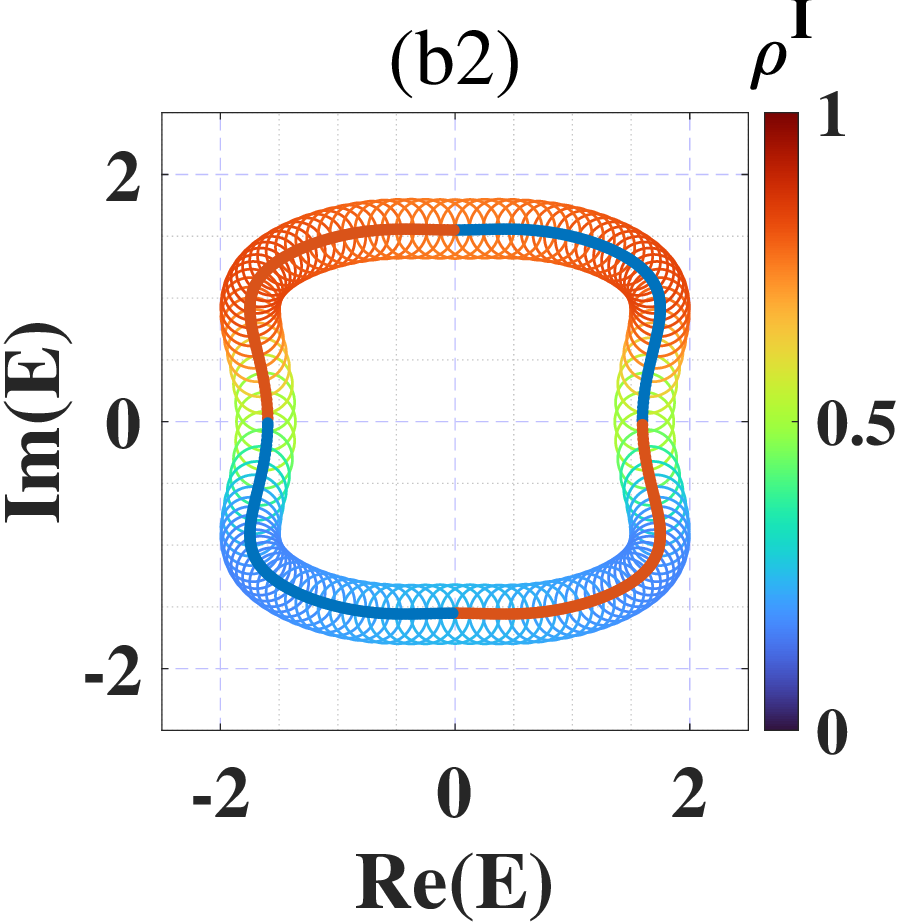}
    \end{minipage}
        \hspace{0.15cm}
    \begin{minipage}[b]{0.18\linewidth}
        \includegraphics[scale=0.235]{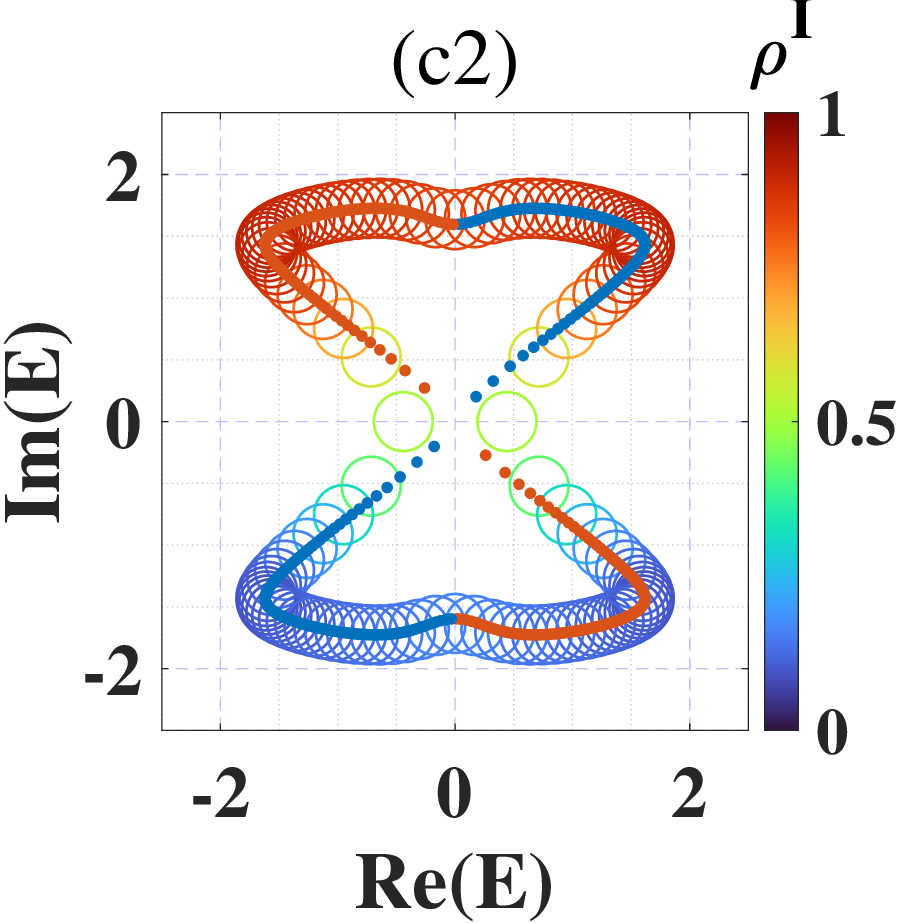}
    \end{minipage}
        \hspace{0.15cm}
    \begin{minipage}[b]{0.18\linewidth}
       \includegraphics[scale=0.235]{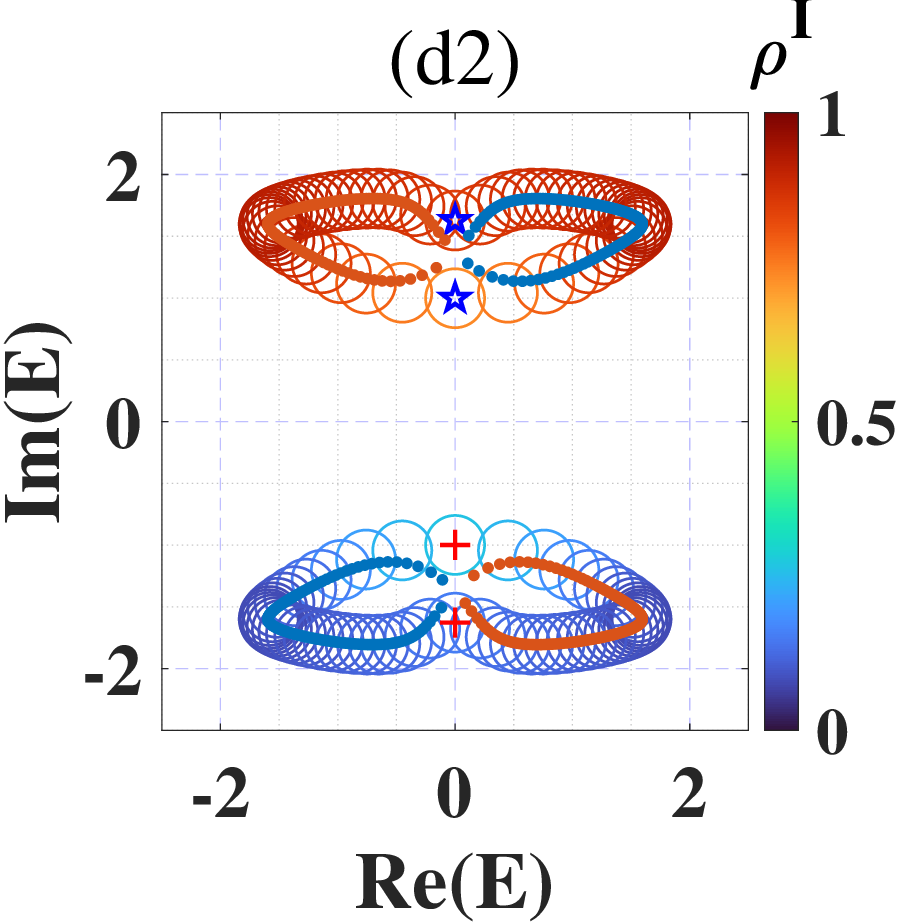}
    \end{minipage}
        \hspace{0.15cm}
    \begin{minipage}[b]{0.18\linewidth}
         \includegraphics[scale=0.24]{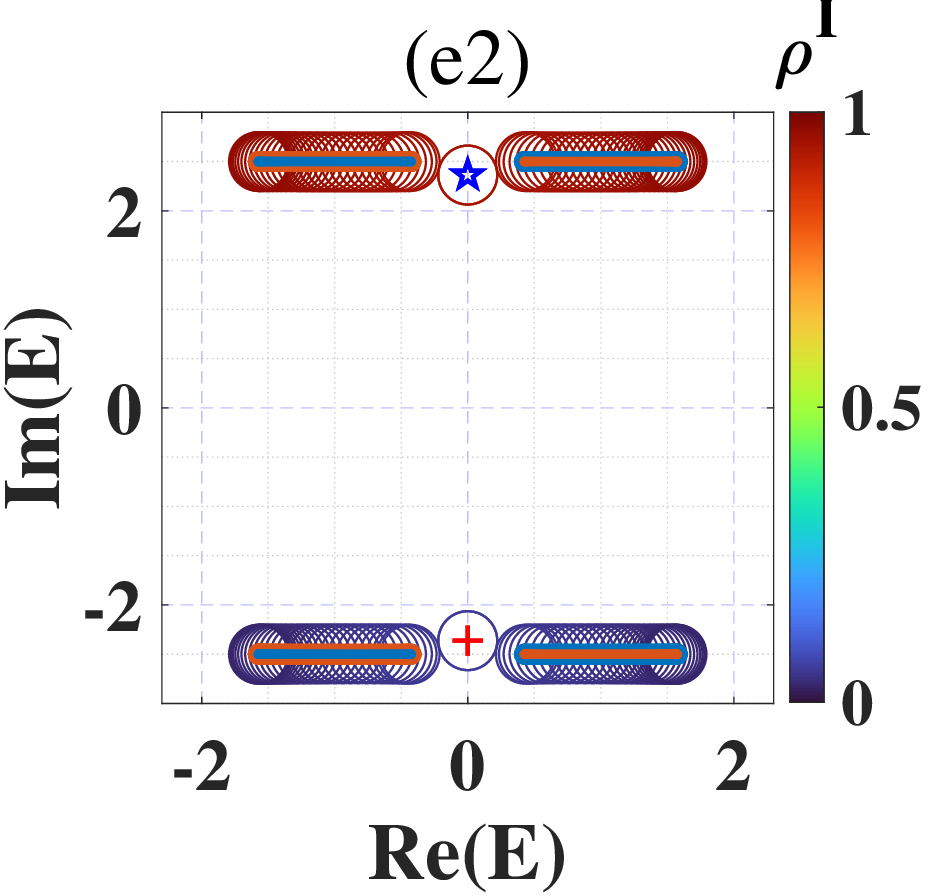}
    \end{minipage}
     \hspace{0.2cm}
    \begin{minipage}[b]{0.18\linewidth}
       \includegraphics[scale=0.22]{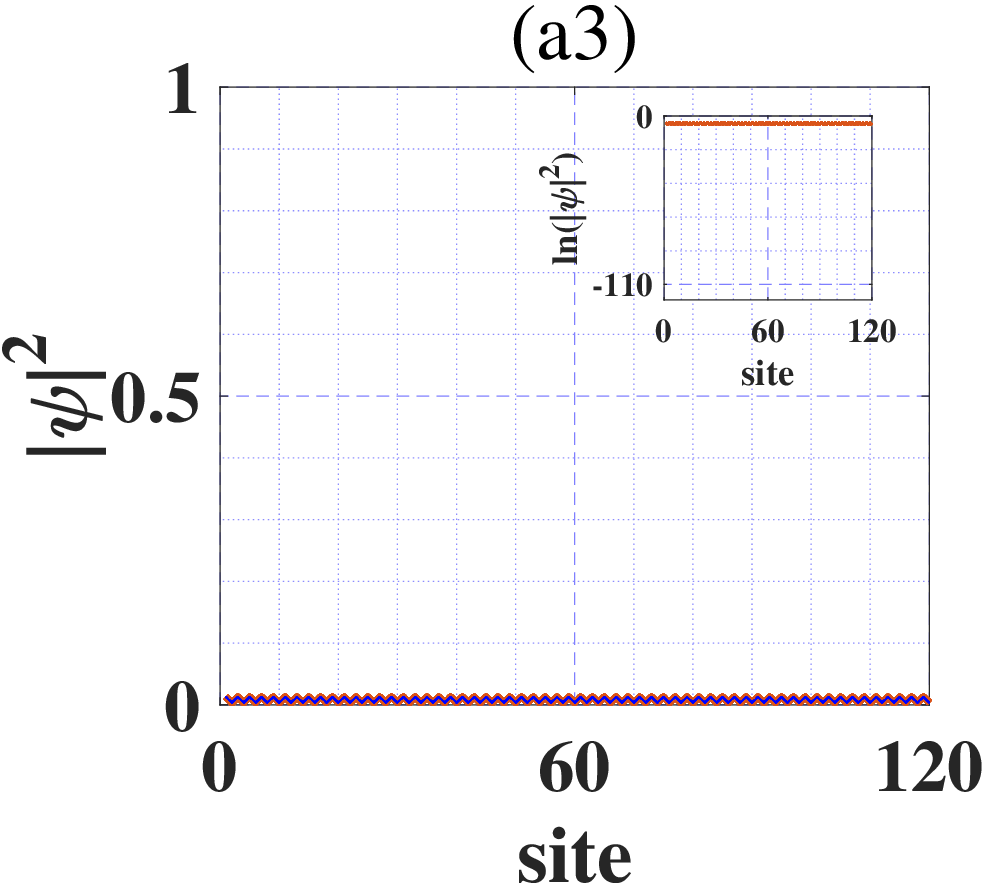}
    \end{minipage}
      \hspace{0.2cm}
    \begin{minipage}[b]{0.18\linewidth}
        \includegraphics[scale=0.22]{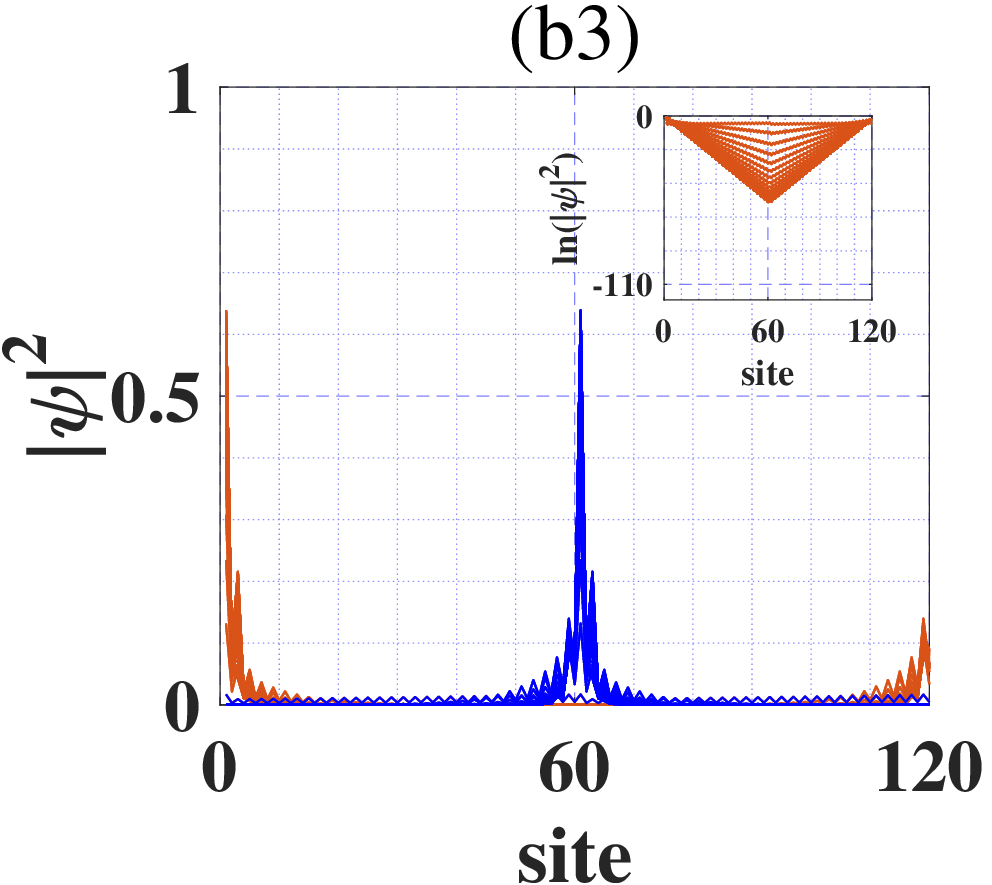}
    \end{minipage}
   \hspace{0.2cm}
    \begin{minipage}[b]{0.18\linewidth}
        \includegraphics[scale=0.22]{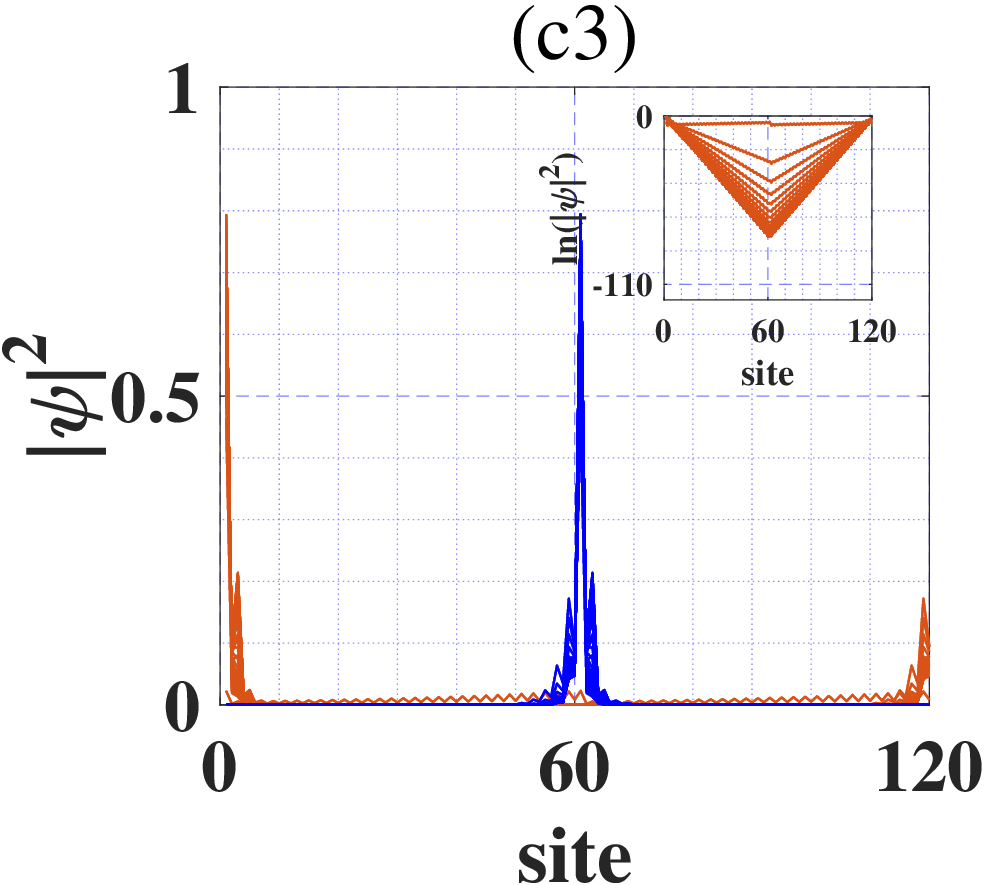}
    \end{minipage}
        \hspace{0.2cm}
    \begin{minipage}[b]{0.18\linewidth}
       \includegraphics[scale=0.22]{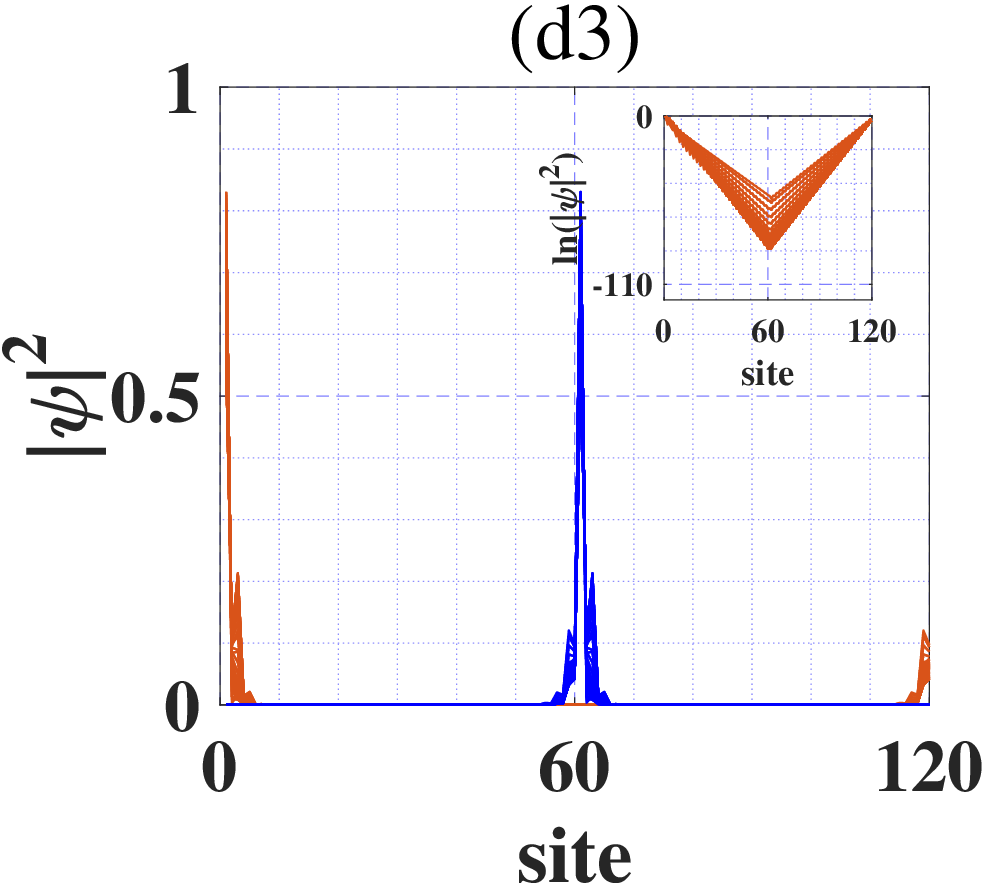}
    \end{minipage}
        \hspace{0.2cm}
    \begin{minipage}[b]{0.18\linewidth}
         \includegraphics[scale=0.22]{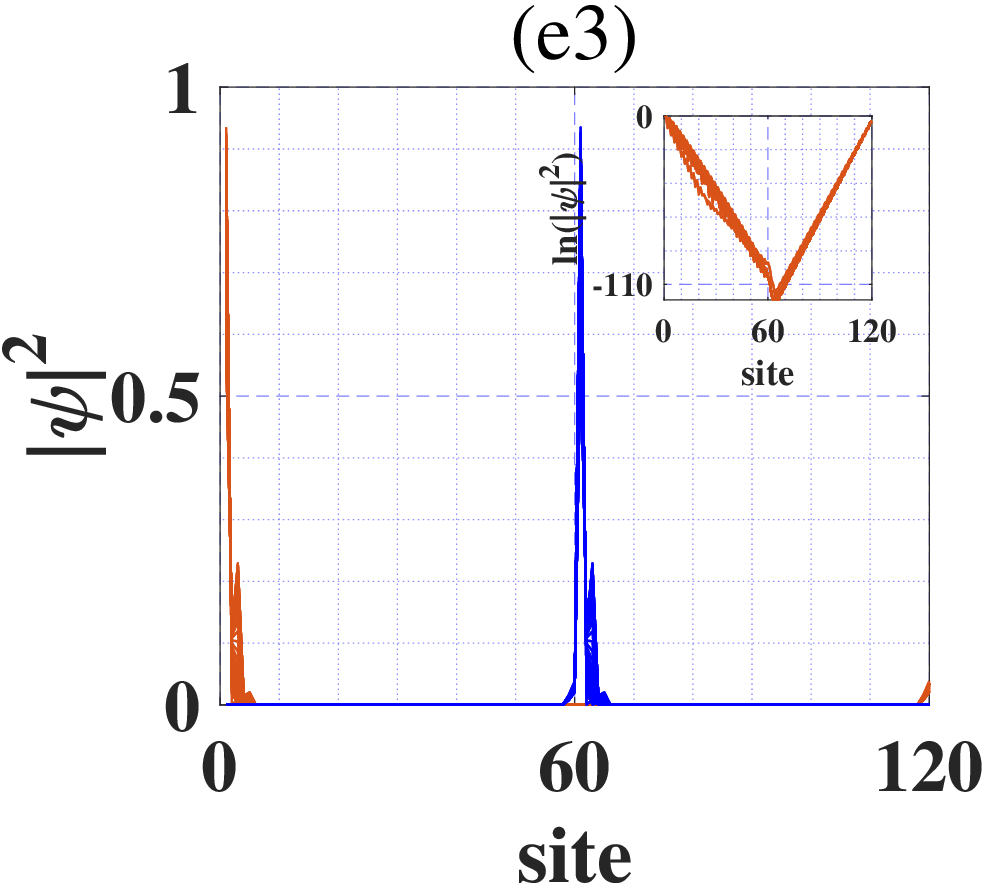}
    \end{minipage}
    \caption{(a1)-(e1) The variation of GBZ with dissipation enhancement, and the black hollow circles indicate the BZ.  (a2)-(e2) Comparison of numerical  (hollow circles) and analytical  (solid circles) energy spectrum. $\rho^{\rm I}= \sum_{n=1}^{N^I} |\psi_n|^2 $ the probability that the eigenstates is localized in chain-${\rm I} $. Blue pentagram and the red cross indicate topological edge states in (d2) and (e2). The color of the analytical energy spectrum corresponds to the GBZ one to one.   (a3)-(e3) The distribution of eigenstates. States with ${\rm Im(E)}>0$ (brown lines) mainly distributed in chain-${\rm I}$. States with ${\rm Im(E)}<0$ (blue lines) mainly distributed in chain-${\rm II}$. The insets in (a3)-(e3) are the wave function after taking the logarithm.  From left to right, $\epsilon=0,1,1.44,1.6$ and $2.5$. Common paratemers: $\rm 2N=60$, $t_1=1.7$, $\gamma=1.6$, $t_2=1$.}  
    \label{fig:4}
\end{figure*}

\begin{figure*}[htbp]
    \centering
    \begin{minipage}[b]{0.18\linewidth}
        \includegraphics[scale=0.21]{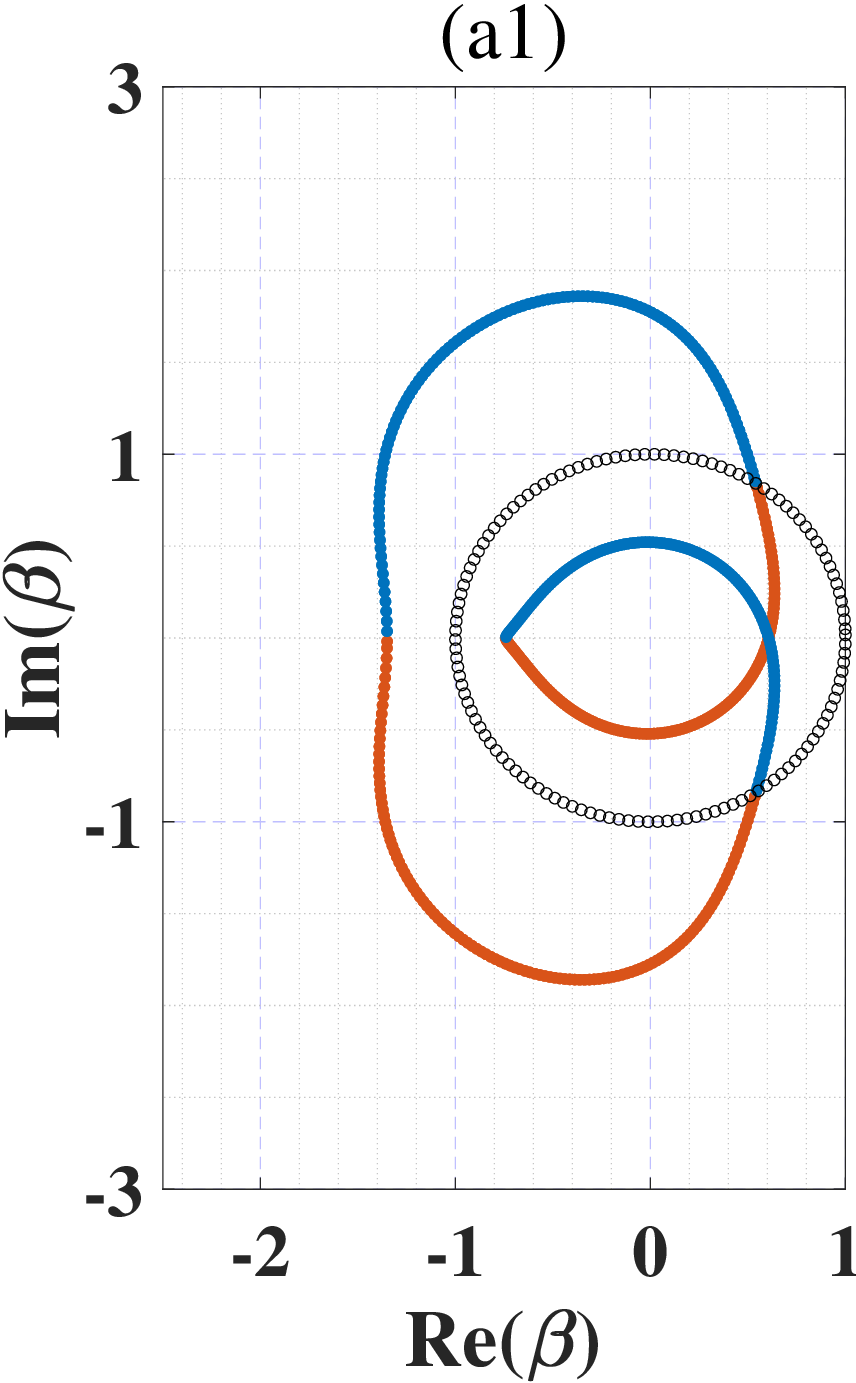}
    \end{minipage}
     \hspace{0.1cm}
    \begin{minipage}[b]{0.18\linewidth}
       \includegraphics[scale=0.22]{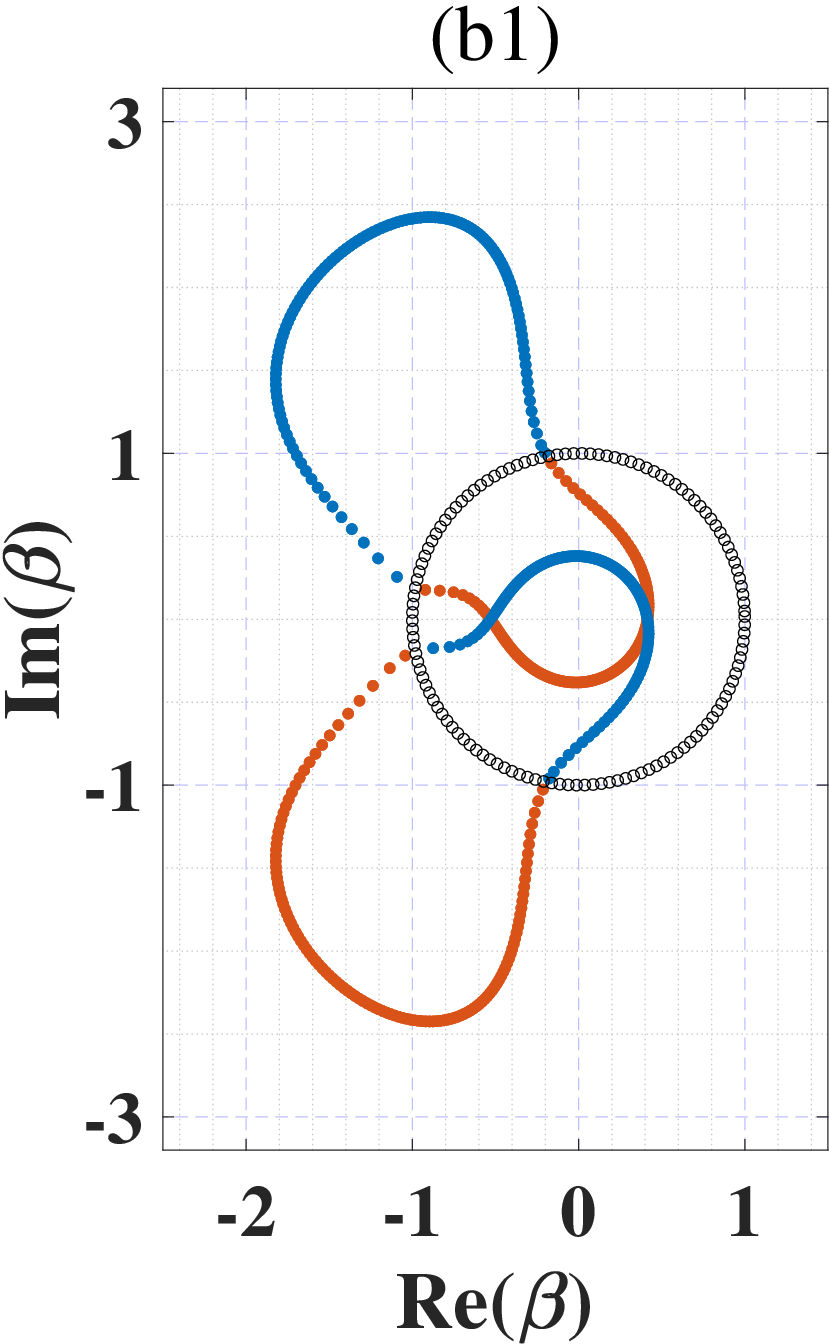}
    \end{minipage}
        \hspace{0.1cm}
    \begin{minipage}[b]{0.18\linewidth}
       \includegraphics[scale=0.22]{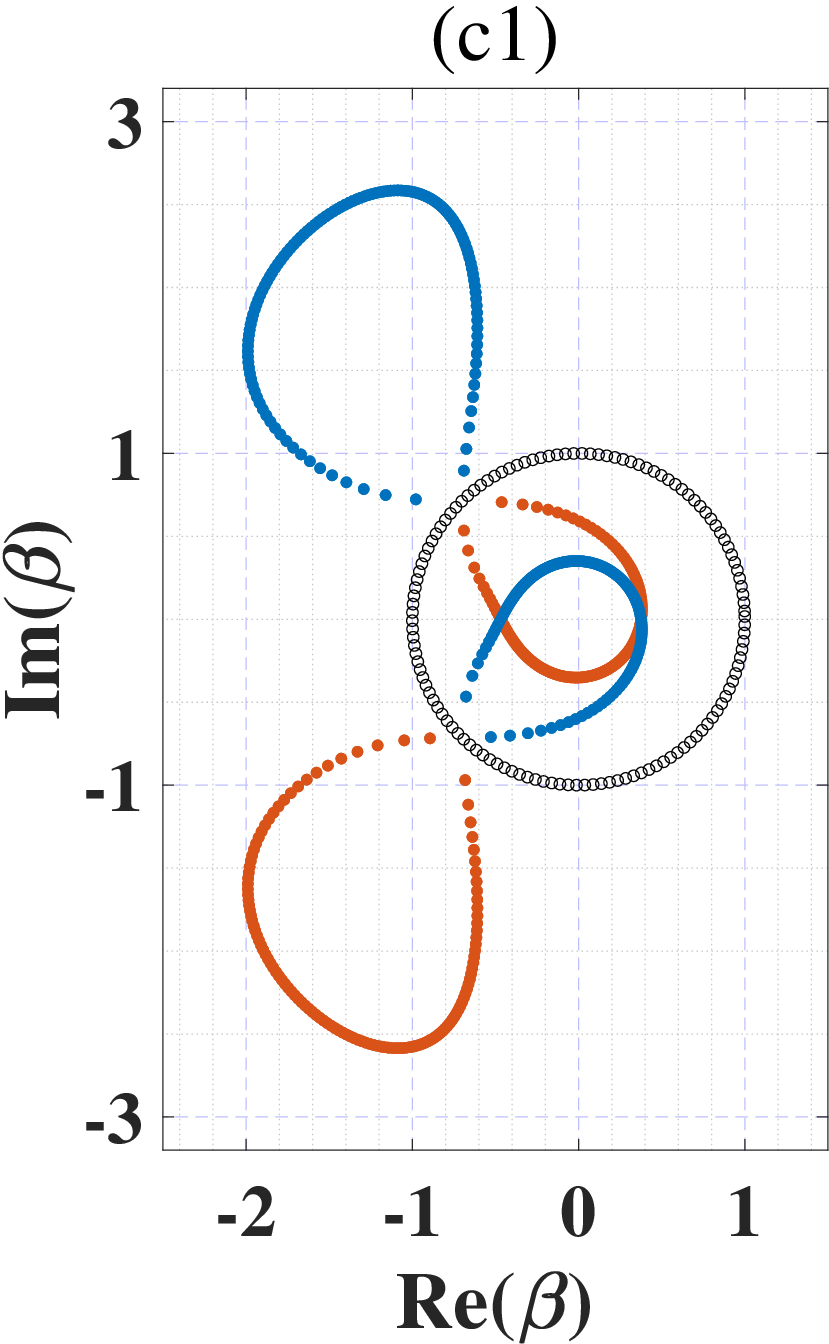}
    \end{minipage}
          \hspace{0.1cm}
    \begin{minipage}[b]{0.18\linewidth}
       \includegraphics[scale=0.22]{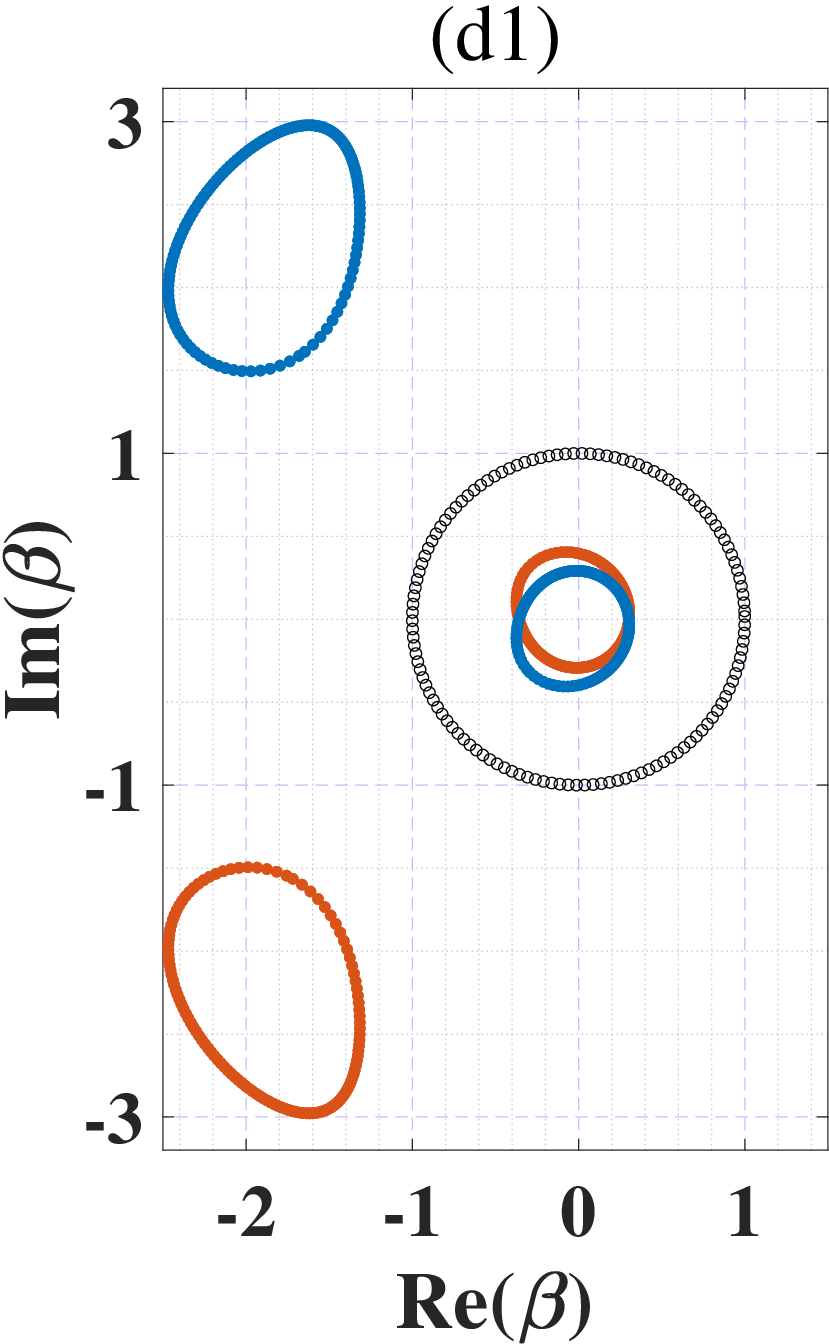}
    \end{minipage}
       \hspace{0.1cm}
    \begin{minipage}[b]{0.18\linewidth}
       \includegraphics[scale=0.22]{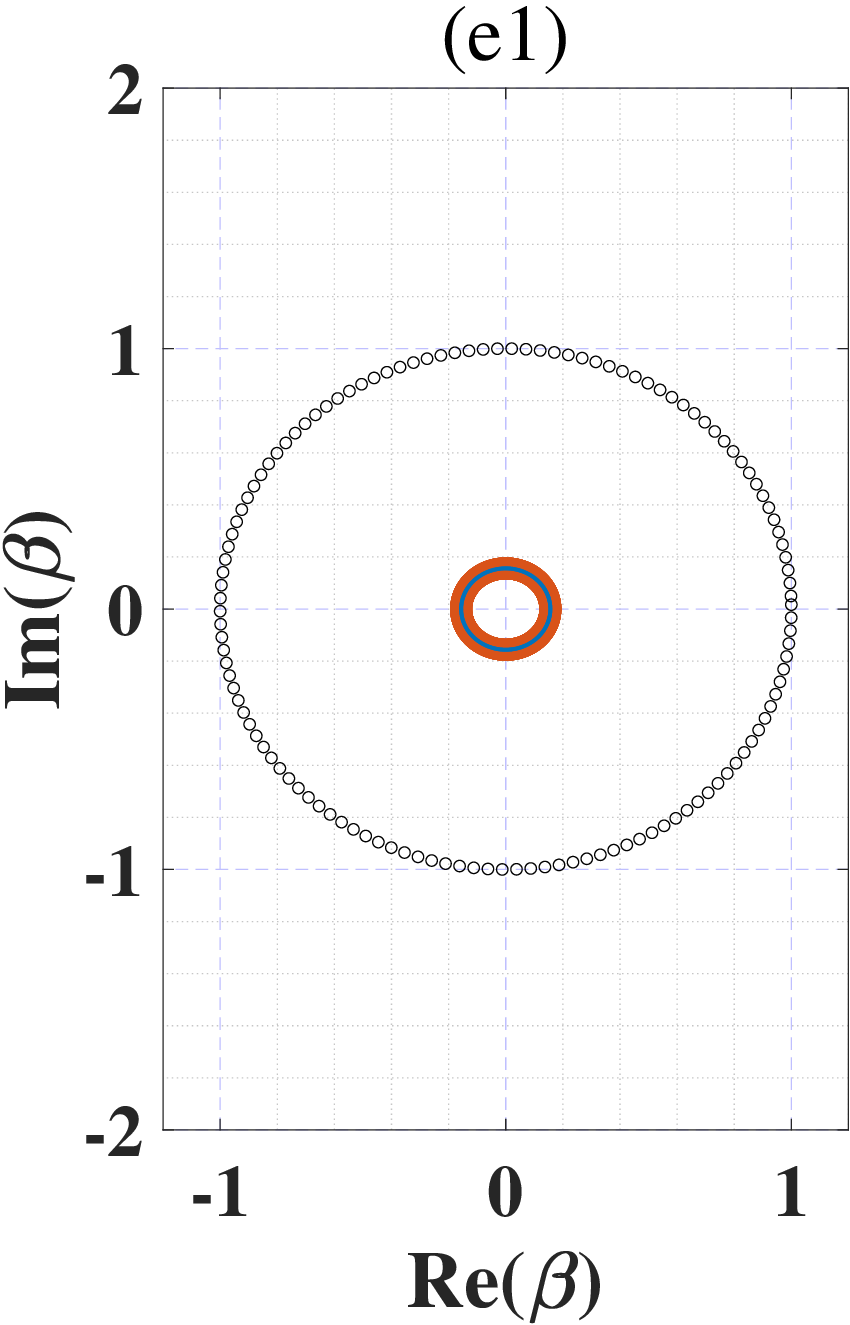}
    \end{minipage}
    \vspace{0.3cm}
    \vspace{0.3cm}
 \begin{minipage}[b]{0.18\linewidth}
       \includegraphics[scale=0.23]{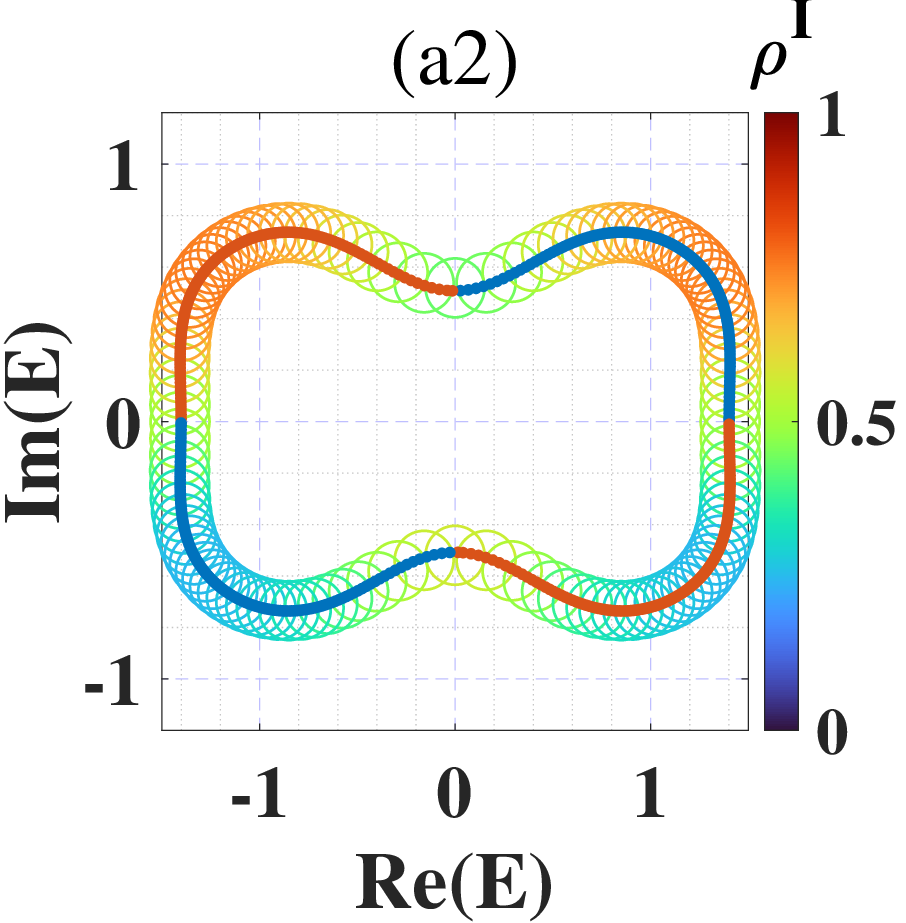}
    \end{minipage}
       \hspace{0.1cm}
    \begin{minipage}[b]{0.18\linewidth}
        \includegraphics[scale=0.23]{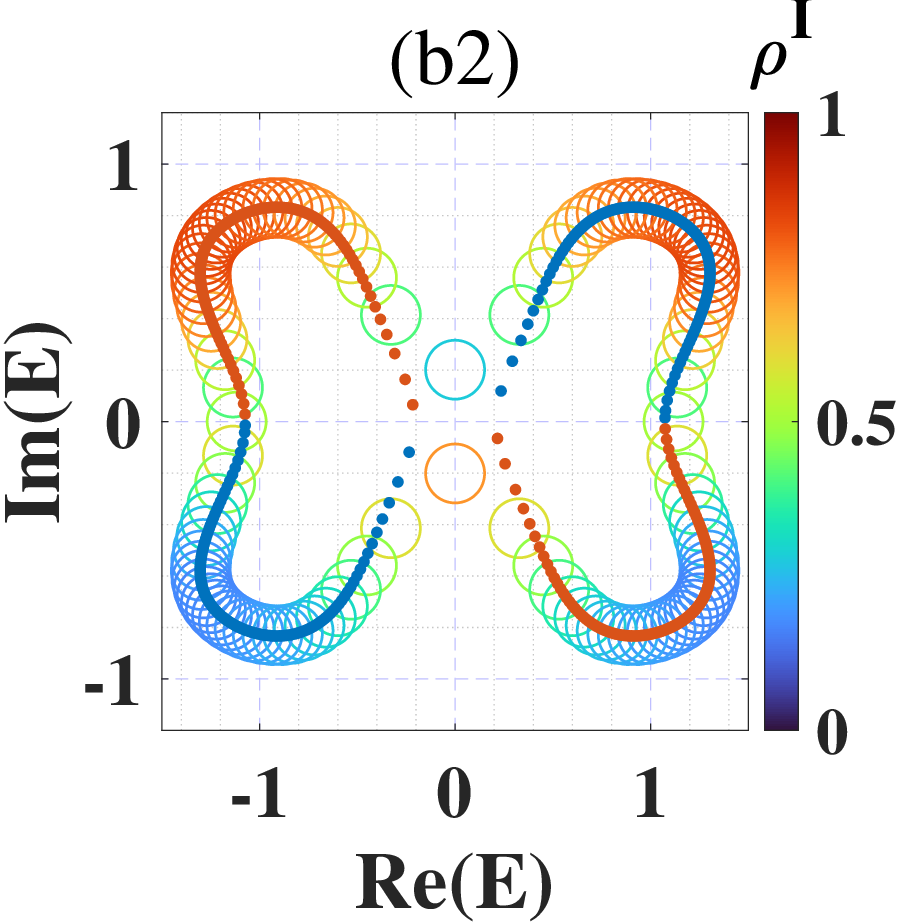}
    \end{minipage}
        \hspace{0.1cm}
    \begin{minipage}[b]{0.18\linewidth}
        \includegraphics[scale=0.23]{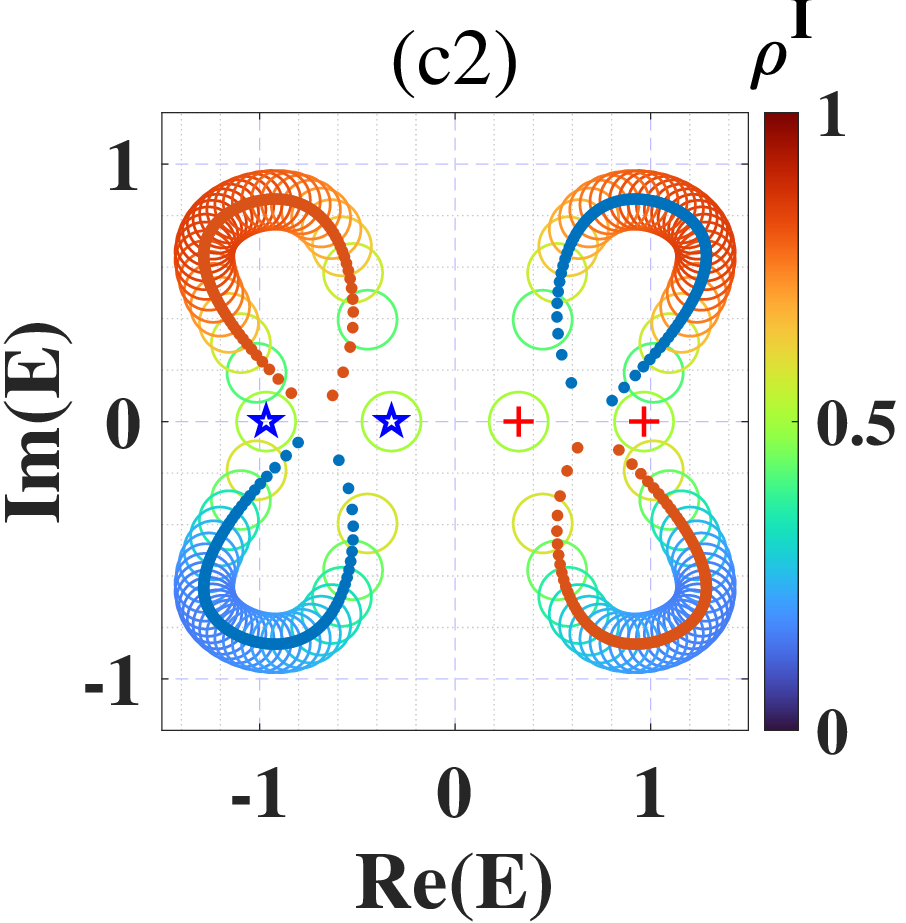}
    \end{minipage}
        \hspace{0.1cm}
    \begin{minipage}[b]{0.18\linewidth}
       \includegraphics[scale=0.23]{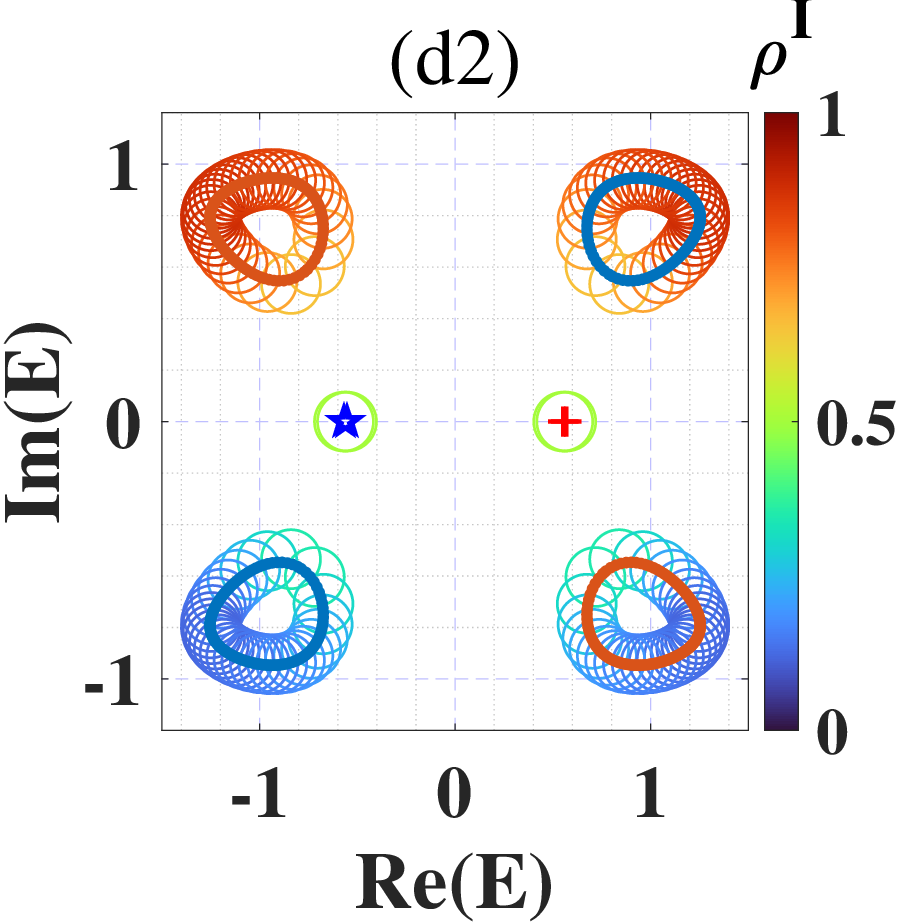}
    \end{minipage}
        \hspace{0.1cm}
    \begin{minipage}[b]{0.18\linewidth}
         \includegraphics[scale=0.236]{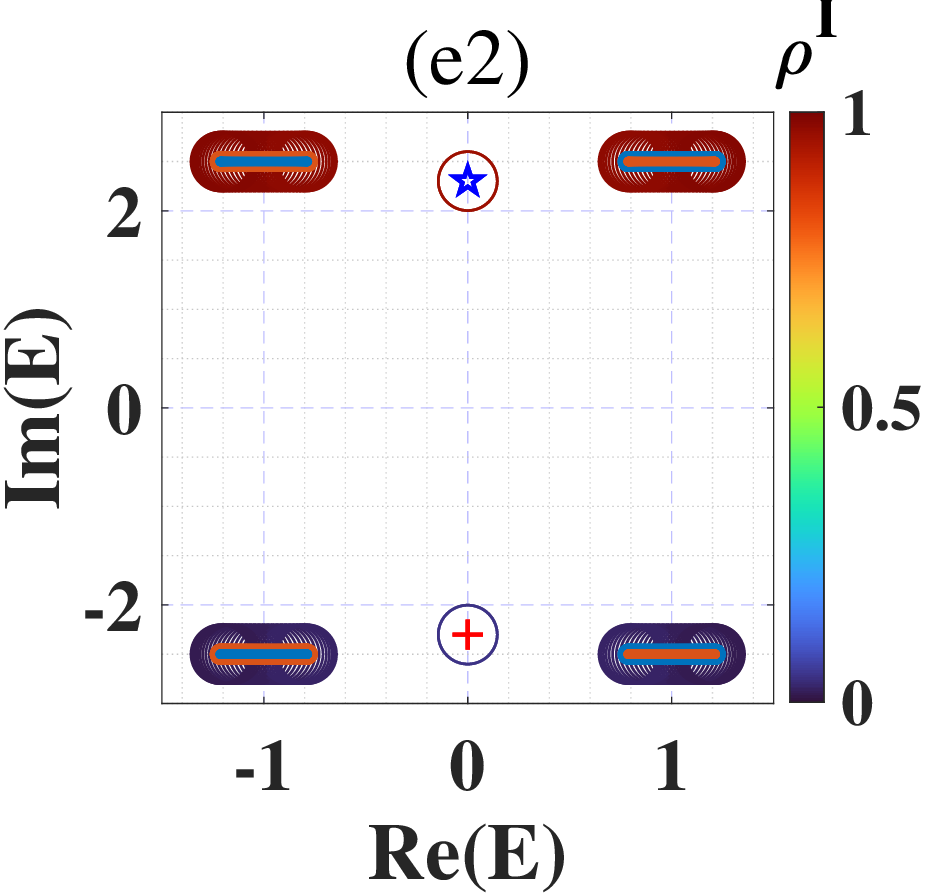}
    \end{minipage}
     \caption{(a1)-(e1) The variation of GBZ with dissipation enhancement, and the black hollow circles indicate the BZ.  (a2)-(e2) Comparison of numerical  (hollow circles) and analytical  (solid circles) energy spectrum. $\rho^{\rm I}$ the probability that the eigenstates is localized in chain-${\rm I} $. Blue pentagram and the red cross indicate topological edge states in (c2) - (e2). The color of the analytical energy spectrum corresponds to the GBZ one to one.  From left to right, $\epsilon=0.4,0.61,2/3,0.8$ and $2.5$. Common paratemers: $\rm 2N=60$, $t_1=0.7$, $\gamma=2/3$, $t_2=1$.}
    \label{fig:5}
\end{figure*}

To enhanced visualization of the individual phases and phase transitions outlined in Sect. \ref{sec:Evolution}, two evolutionary processes of the spectral structures are depict in Figs. \ref{fig:4} and \ref{fig:5}, corresponding respectively to points (a)-(e) in Figs. \ref{fig:3}(A) and (B). The numerical (hollow circle) and analytical (solid circle) energy spectra are in excellent agreement, as shown in Figs. \ref{fig:4}(a2)-(e2) and Figs. \ref{fig:5}(a2)-(e2).
In the absence of dissipation, the system maintains translational invariance, exhibiting a closed-loop structure in the energy spectrum (Fig. \ref{fig:4}(a2)). Although the non-reciprocal SSH model has two bands due to its double sublattice structure (as shown in Eq.(\ref{eq:3})), these bands are connected under PBC for parameters in the topological region, forming a closed ring structure \cite{Xiong2018Why}. At this stage, the eigenstates are extended (Fig. \ref{fig:4}(a3)), uniformly distributed along the entire chain, which is also evident from the colorbar on the numerical energy spectrum (Fig. \ref{fig:4}(a2)). The corresponding $C_{\beta}$ coincides with the BZ (Fig. \ref{fig:4}(a1)), indicating the absence of a skin effect.

When the dissipation is weak, the system tends to maintain translational invariance, so the energy spectrum still exhibit a closed-loop structure (Fig. \ref{fig:4}(b2) and Fig. \ref{fig:5}(a2)), similar to that of non-reciprocal systems with translational invariance. Under the tearing effect \cite{2024tearing} and non-reciprocal interaction, the energy spectrum is partially torn, meaning that two inconspicuous boundaries are formed at the domain wall, then states with $\rm Im(E) > 0$ are mainly distributed at the left boundary of chain I (brown lines), while states with $\rm Im(E) < 0$ are primarily located at the left boundary of chain II (blue lines), giving rise to the skin effect (as shown in Fig. \ref{fig:4}(b3)).
Interestingly, a small number of states are also distributed at the right boundaries of chains I and II, which is different from the traditional skin effect induced by nonreciprocity.  This occurs because the dissipation is weak, allowing the eigenstates localized at the left boundaries of chains I and II to jump over the domain wall and enter the right boundaries of chains II and I, respectively. This creates what appears to be a rightward skin effect, which we refer to as the residual effect. However, this phenomenon is not actually a skin effect. The skin effect is typically due to non-reciprocal interaction, but in this case, the non-reciprocal coupling to the left is stronger. We can conclude that each state is composed of two parts: the skin state and the residual state, which are located on the two sides of the domain wall.

Eigenstates with larger $| \rm E|$ are more conspicuously torn according to colorbar, indicating a stronger localized skin effect, a smaller localization length, and a weaker residual effect. Consequently, the localization lengths of eigenstates with different $| \rm E|$ vary, a difference that can be observed through the logarithm of the wave functions, as it characterizes the localization length. These varying localization lengths are also reflected by the GBZ curve $C_{\beta}$, given the one-to-one correspondence between the localization length and the mode length of $C_{\beta}$ ($\beta=\sqrt{e^{1/a}}$ with $a$ being the localization length numerically) \cite{PhysRevLett.121.086803}. As shown in Fig. \ref{fig:4}(b1) and Fig. \ref{fig:5}(a1), under weak dissipation, the $C_{\beta}$ deviates from the BZ, forming a double surround structure both inside and outside the BZ. The $C_{\beta}$ inside the BZ corresponds to the skin states localized at the left boundaries of the two chains, while the $C_{\beta}$ outside the BZ corresponds to the residual states localized at the right boundaries of the two chains. Together, the skin and residual states on either side of each domain wall constitute a complete state. In other words, each point on the energy spectrum corresponds to two points of the same color on the GBZ, both inside and outside the BZ, with the colors of the energy spectrum matching one-to-one with those of the $C_{\beta}$. 

As the dissipation increases, the energy spectrum is translated along the positive and negative directions of the imaginary axis due to the tearing effect of the dissipation domain wall \cite{2024tearing}. When an Exceptional Point (EP) emerges, an imaginary gap opens, as shown in Fig. \ref{fig:4}(c2) and Fig. \ref{fig:5}(c2). The corresponding  $C_{\beta}$ breaks at the BZ ($ |\beta|=1$), causing $C_{\beta}$ to separate into distinct curves inside and outside the BZ, as illustrated in Fig. \ref{fig:4}(c1) and Fig. \ref{fig:5}(c1). The energy spectrum also shifts along the positive and negative directions of the real axis. This shift is attributed to the double sublattice structure of the SSH model, which results from its double-band nature. After the EP is reached, a real gap opens, as depicted in Fig. \ref{fig:4}(d2) and Fig. \ref{fig:5}(b2).  The corresponding $C_{\beta}$  inside and outside the BZ break and recombine at $\rm Im(\beta)=0$ (Fig. \ref{fig:4}(d1)), forming independent closed curves. Alternatively, a complete double loop structure may break and recombine to form two separate closed curves, as shown in Fig. \ref{fig:5}(b1).

Regardless of whether the imaginary gap opens first followed by the real gap (Fig. \ref{fig:4}(a2)-(e2)), or the real gap opens first followed by the imaginary gap (Fig. \ref{fig:5}(a2)-(e2)), the key outcome manifests as concurrent real and imaginary band gap openings, which divide the energy spectrum into four parts and is accompanied by the emergence of topological edge states (denoted by blue pentagrams and red crosses) --- a phenomenon we term 'complete tearing'. Interestingly, the opening of the real gap, which is caused by the transition of the system from PBC to OBC, mirrors observations made in previous study \cite{Xiong2018Why}, in which a similar effect was noted when boundary hopping amplitude is slowly eliminate to reach OBC. During the tearing of the energy spectrum, the corresponding $C_{\beta}$ curves gradually break and reassemble both inside and outside the BZ. This process results in the formation of four ring structures, both inside and outside the BZ, as shown in Fig. \ref{fig:4}(b1)-(d1) and Fig. \ref{fig:5}(a1)-(d1). This indicates that the localization lengths of all bulk states tend to become uniform, which can also be intuitively seen in Fig. \ref{fig:4}(b3)-(d3). Under traditional OBC, bulk states in nonreciprocal systems typically localize to a boundary. However, in our model, the dissipation domain wall replaces the traditional OBC. It drives gradually a skin effect induced by nonreciprocal hopping, and a small fraction of these states jump over the domain wall, localizing at the opposite side of the domain wall, resulting in a type of skin effect intermediate between the extended states under PBC and the localized states under OBC. It's worth noting that the opening of the imaginary gap consistently corresponds to the separation of $C_{\beta}$ inside and outside the BZ. This is because the dissipation added to the two chains is symmetric, causing the imaginary gap to open at $\rm Im(E)=0$. States with $\rm Im(E)=0$ are minimally affected by dissipation and thus remain extended, aligning precisely with the bulk states described by the BZ. While the opening of the real gap is always associated with a double surround structure of the $C_{\beta}$  curve breaking and recombining to form two closed curves. 

\begin{figure}[htbp]
\begin{center}
\includegraphics[scale=0.5]{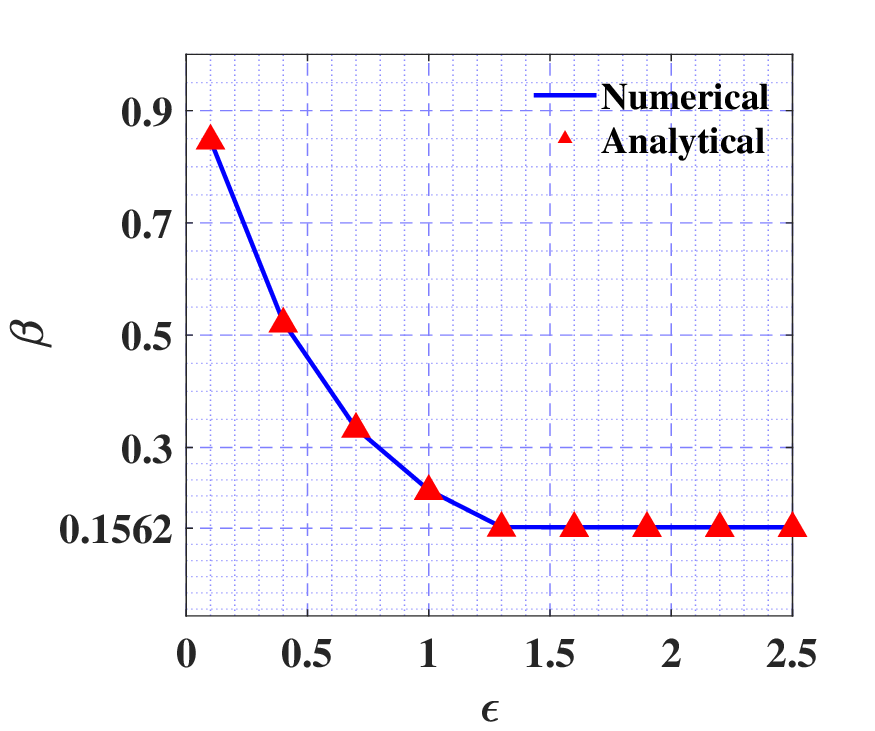}
\caption{The local length of the eigenstates with the largest  $|\rm E|$  varies with dissipation enhancement for $t_{1}=0.7$, $\gamma=2/3$.  Blue solid lines indicate numerical results, and red triangles indicate  analytical results. }
\label{fig:6}
\end{center}
\end{figure}

When the dissipation is further enhanced, a system with PBC will be effectively torn into two systems with OBC. This can be observed from the energy spectrum, the eigenstates, their localization lengths, and the $C_{\beta}$. The energy spectrum gradually contracts from four rings into four arcs, resembling the structure under OBC, as shown in Fig. \ref{fig:4}(e2) and Fig. \ref{fig:5}(e2). States with $\rm Im(E) > 0$ are almost entirely localized in chain I, while states with $\rm Im(E) <0$ are almost entirely localized in chain II, as depicted in Fig. \ref{fig:4}(e3). Meanwhile, the logarithm of the wave function also becomes nearly identical, and the localization length ultimately matches that under OBC, as shown in Fig. \ref{fig:6}. The corresponding $C_{\beta}$ exists only within the BZ, similar to the situation under OBC, as illustrated in Fig. \ref{fig:4}(e1) and Fig. \ref{fig:5}(e1). This is because the $C_{\beta}$ outside the BZ describes residual  states, which are localized at the right boundary of the two chains. The disappearance of the $C_{\beta}$ outside the BZ indicates that no states are jumping across the domain wall, meaning that the two chains have become independent of each other.

Fig. \ref{fig:6} illustrates the variation of the numerical and analytical localization lengths with increasing dissipation. To facilitate comparison, a transformation was applied to the numerical localization length: $a \rightarrow \beta = \sqrt{e^{1/a}}$. The figure shows that the numerical and analytical results are in excellent agreement, thereby validating the accuracy of the GBZ theory.
Moreover, when the dissipation is sufficiently large, both the numerical and analytical localization lengths tend towards a stable value. This stable value precisely corresponds to the GBZ of the non-reciprocal SSH model under OBC: $\beta=\sqrt{(t_1+\gamma)/(t_1-\gamma)}=0.1562$. This further demonstrates that a chain with PBC is effectively torn into two independent chains with OBC.
 
As the dissipation increases, the energy spectrum evolves through a series of distinct phase. Initially, it features a single ring structure with translation invariance. Gradually, this structure expands in four directions, eventually splitting into four smaller rings. Ultimately, these rings contract into four arcs. In this process, the localization degree of the various eigenstates becomes increasingly uniform, with their localization lengths ultimately converging to those observed under OBC. The corresponding $C_{\beta}$ curve undergoes significant changes as well. Initially, it breaks and reassembles both inside and outside the BZ. Eventually, it stabilizes and exists solely within the BZ. Although the system maintains PBC throughout, the tearing effect of the dissipation domain wall causes the energy spectrum, eigenstates, localization lengths, and $C_{\beta}$ to exhibit the same structure and characteristics as those under OBC. This process illustrates the transition of the system from PBC to OBC and highlights the continuous nature of the breakdown of the BBC in the gain and loss domain wall system.

\subsection{The emergence of topological edge states}\label{subsec:IVC}
\begin{figure*}[ht]
\begin{center}
\hspace{-1cm}
\includegraphics[scale=0.23]{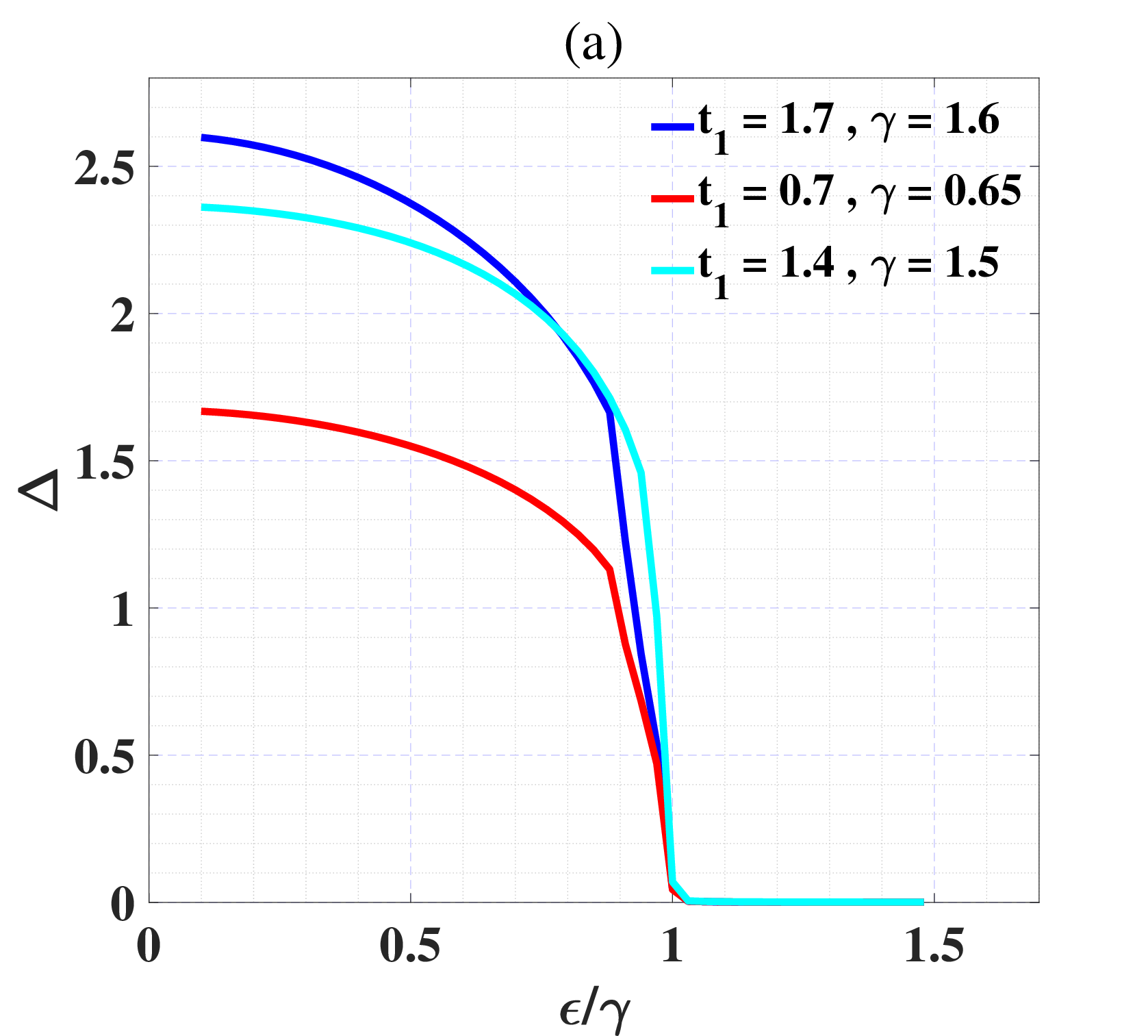}
\includegraphics[scale=0.25]{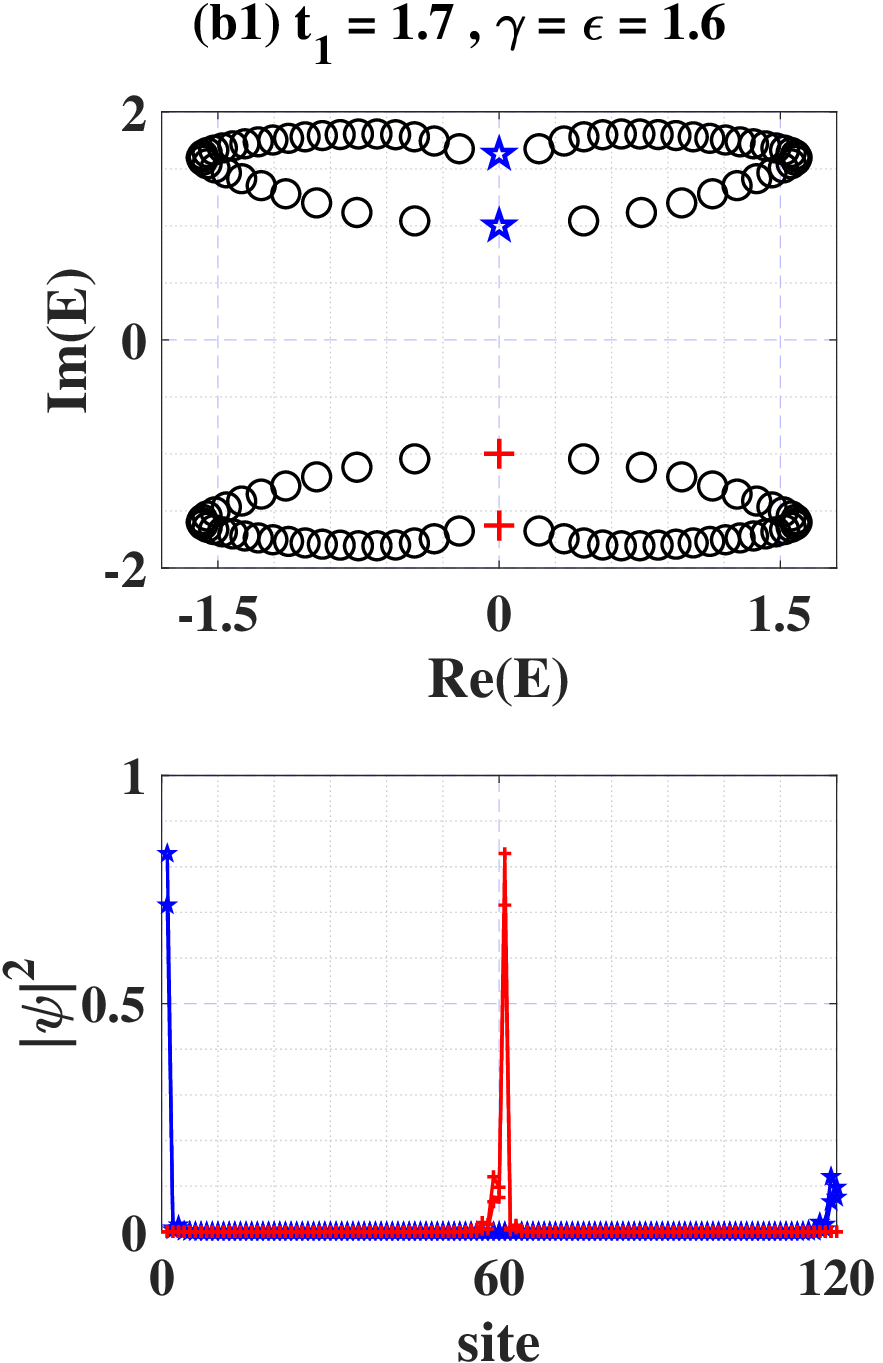}
\includegraphics[scale=0.25]{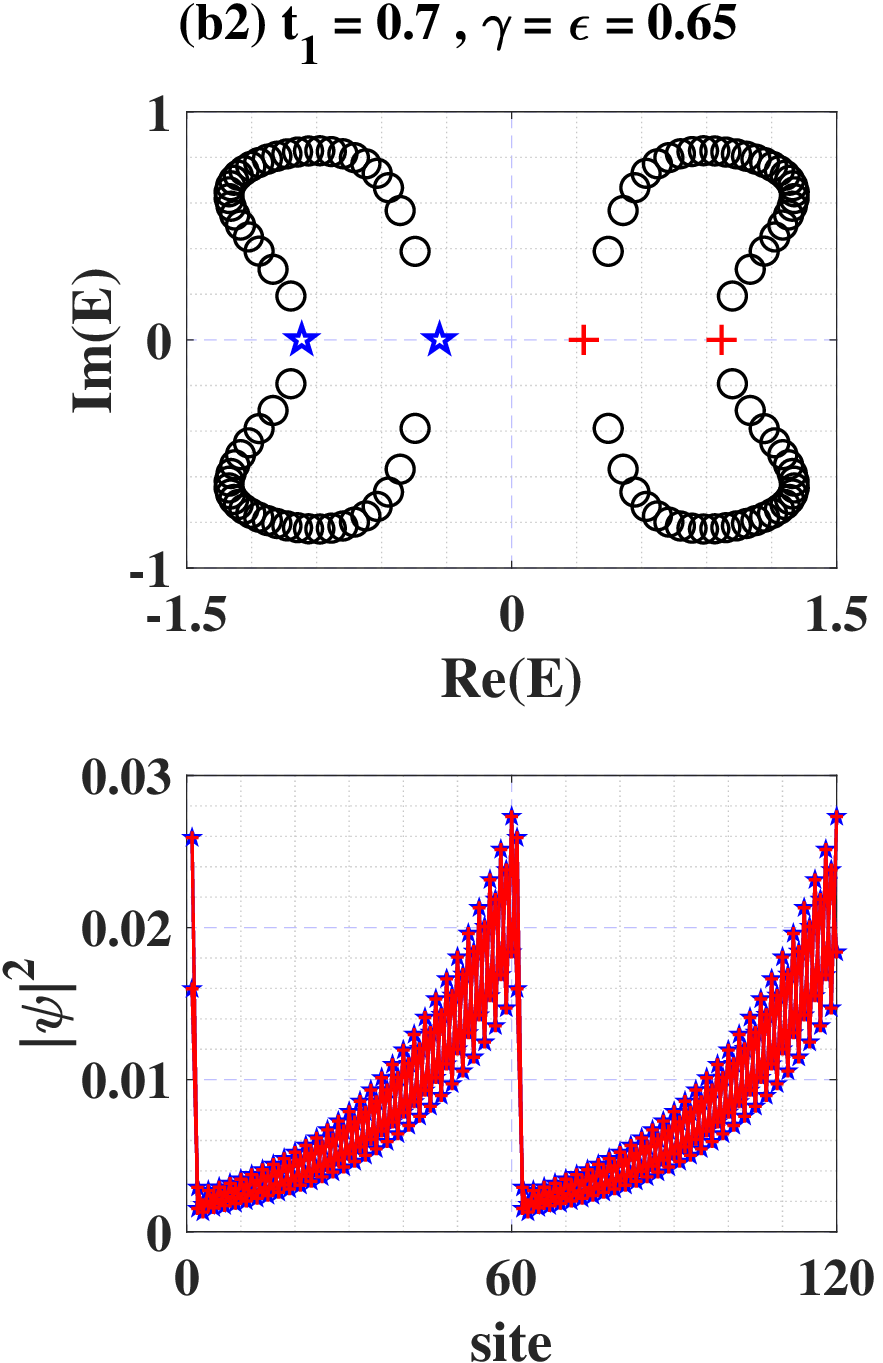}
\includegraphics[scale=0.25]{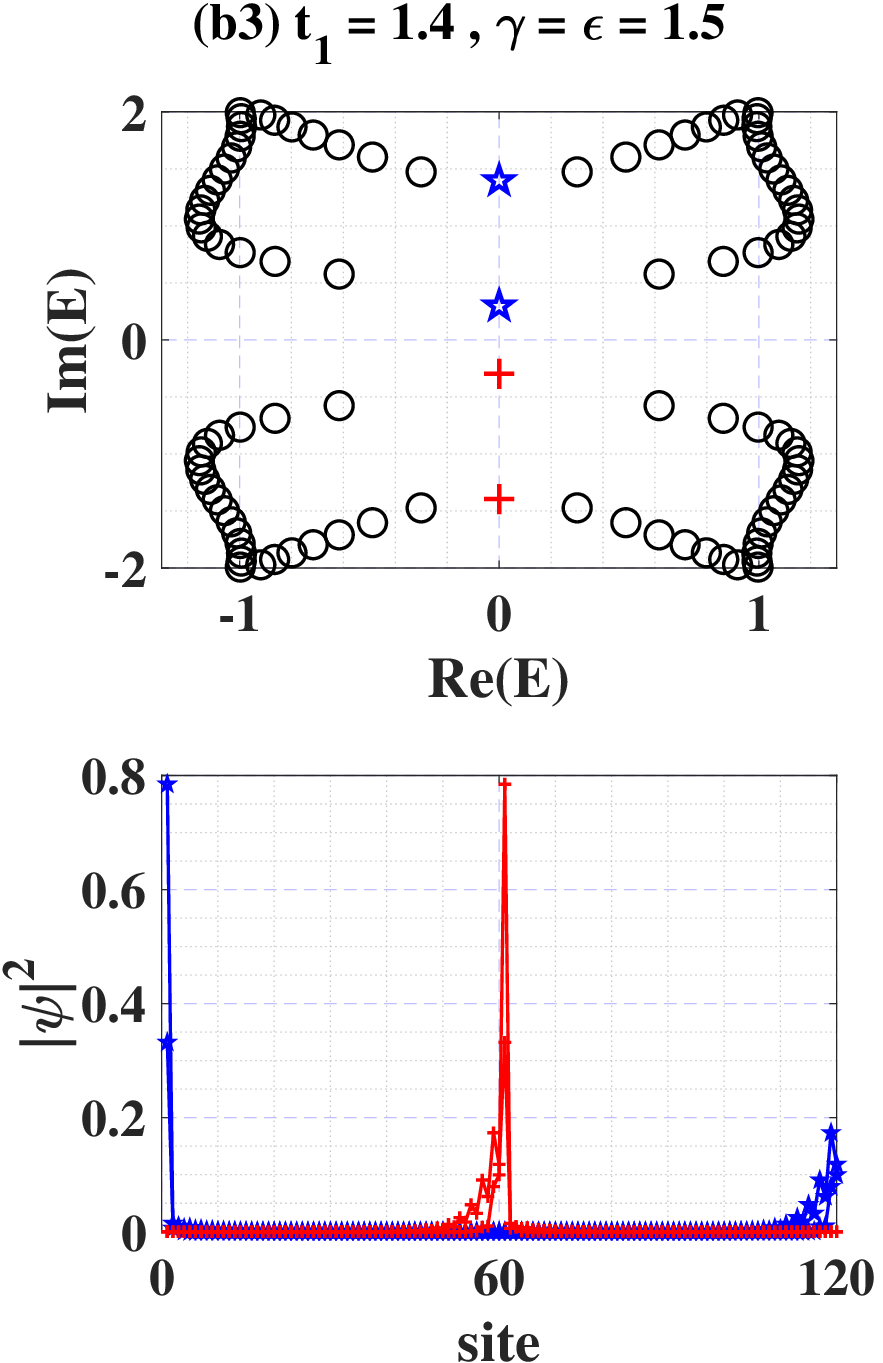}
\caption{(a) The $\Delta$  varies as $\epsilon/{\gamma}$ for different parameters. (b1)-(b3) The energy spectrum and the distribution of topological edge states for the corresponding parameters, and both the blue pentagram and the red cross indicate topological edge states.}
\label{fig:7}
\end{center}
\end{figure*}
The topological region under iPBC is distinct from both under PBC and OBC. It is closely related to the dissipation strength, and gradually expands along the topological phase boundary ($t_1=\pm \sqrt{t_2^2+\gamma^2}$) under OBC as the dissipation increases. From Sect. \ref{subsec:IVB}, we know that when parameters are taken from the topological region under OBC,  the complete tearing of the energy spectrum is accompanied by the emergence of topological edge states, that is, the critical point where the spectrum is split into four parts, and the corresponding analytical energy spectrum forms a loop in each quadrant. Subsequently, under the same parameters, the structure and properties of the energy spectrum, eigenstates, and the GBZ gradually transition to those under OBC as dissipation further increases. To determine the emergence of topological edge states, it is essential to quantify the critical dissipation strength required for the energy spectrum to be completely torn.

In the process of solving the GBZ equation, as  $\rm \theta$ in Eqs.(\ref{D-7}) and (\ref{D-8}) varies from $0$ to $2 \pi$, the analytical energy in each quadrant rotates counterclockwise. Eventually, these energies connect end-to-end to form a closed curve, as shown in Fig. \ref{fig:4}(d2) and Fig. \ref{fig:5}(c2). Simultaneously, the corresponding $C_{\beta}$ curves also rotate counterclockwise both inside and outside the BZ, as depicted in Fig. \ref{fig:4}(d1) and Fig. \ref{fig:5}(c1). This indicates that the energy spectrum has been completely torn. Thus, this end-to-end connection point marks the phase transition where complete tearing of the energy spectrum and the emergence of topological edge states occur. We define $\Delta=\rm E(0^{+})-E(0^{-})$, where $ \rm E(0^{+}) $ and $\rm E(0^{-})$   represent the energies calculated when $\rm \theta$ is close to $0$ to $2 \pi$, respectively. Consequently, $\Delta=0$  signifies that the analytical energy in this quadrant is connected end-to-end. The corresponding $\epsilon$ value thus represents the critical dissipation strength required for the complete tearing of the energy spectrum and the emergence of topological edge states.

Fig. \ref{fig:7} illustrates the variation of $\Delta$ as a function of the ratio of $\epsilon$ to $\gamma$, along with the energy spectrum and the distribution of topological edge states under different parameters. These parameters are taken from the topological region under OBC. As shown in Fig. \ref{fig:7}(a), the slope of $\Delta$ abruptly changes to zero at $\epsilon=\gamma$ for various parameters, indicating that the system is completely torn, that is, when $\epsilon$ and $\gamma$ are nearly equal, the system undergoes a second  EP. The case $\epsilon=-\gamma$ is omitted because it has the same effect. This sudden change can be understood as follows: the non-reciprocal coupling satisfies translational invariance in a system with PBC, however, dissipation breaks this translational invariance. Consequently, the system is completely torn when $\epsilon$  and $\gamma$ are of comparable strength. Figs. \ref{fig:7}(b1)-(b3) depict the numerical energy spectrum and the distribution of topological edge states under the corresponding parameters. It can be seen that the energy spectrum is completely torn at $\epsilon=\gamma$ under different parameters,  being divided into four parts, accompanied by the emergence of topological boundary states (indicated by blue pentagrams and red crosses), then the topological edge states persist for $|\gamma|<\epsilon$. Thus, under iPBC, the phase boundary of the topological region is determined by two conditions: 1. $t_1=\pm \sqrt{t_2^2+\gamma^2}$, 2. $\epsilon=\pm \gamma$.

\section{conclusion}\label{sec:conclu}

We investigate how BBC breaks down in 1D gain-loss domain wall system. As the dissipation is enhanced, system gradually evolves from the PBC to the OBC, then one system with PBC is torn into two systems with OBC ultimately, which can be convincingly demonstrated by the excellent agreement between numerical and analytical results. In this process, the phase diagrams exhibit the hybrids of those under PBC and OBC, whose topological regions are both dissipation-tunable and analytically solvable. This phase diagram hybridization process manifests the continuous nature of BBC breaks down, contrasting sharply with the conventional abrupt BBC failure caused by infinitesimal boundary hopping amplitudes in the thermodynamic limit. Remarkably, during the BBC breaks down process,  various phases and phase transitions emerge, and each phase transitions corresponding to unique behavioral signatures in the GBZ, revealing rich GBZ configurational. Our work help researcher better understand  non-Hermitian topological systems.

\section*{ACKNOWLEDGMENTS}
We are grateful to Xiao-Ran Wang, Xian-Qi Tong and Zheng Wei  for their valuable suggestions on the manuscript. This work is supported by the National Natural Science Foundation of China (Grant No.12174030) and the National Key Research and Development Program of China (Grant No.2023YFA1406704). 

\appendix
\onecolumngrid
\section{Derivation of The Characteristic Equation}\label{app:A}
 The Schrodinger equation for our model (Eq. (\ref{eq:1})) in real space can be written as :
  \begin{equation} \label{A-1}
	\begin{bmatrix}
	 {\rm i}\epsilon    &t_1+\gamma  &        &           &            &           &           &       &            &t_2  \\
     t_1-\gamma  &{\rm i}\epsilon    &t_2     &           &            &           &           &       &            &  \\
      &          \cdots      &\cdots  &\cdots     &            &           &           &       &            &  \\
      &          &            t_2     &{\rm i}\epsilon   &t_1+\gamma  &           &           &       &            &  \\
      &          &            &        t_1-\gamma &{\rm i}\epsilon    &t_2        &           &       &            &  \\
      &          &            &        &           t_2         &{\rm -i}\epsilon  &t_1+\gamma &       &            &  \\
      &          &            &        &           &            t_1-\gamma &{\rm -i}\epsilon  &t_2    &            &  \\
      &          &            &        &           &            &           \cdots     &\cdots &\cdots      & \\
      &          &            &        &           &            &           &           t_2    &{\rm -i}\epsilon   &t_1+\gamma \\
      t_2          &            &        &           &            &           &          &     &       t_1-\gamma  &{\rm -i}\epsilon
	 \end{bmatrix}
     \begin{bmatrix}
       \Psi^{\rm I}_{1 A} \\
       \Psi^{\rm I}_{1 B}  \\
       \cdots \\
       \Psi^{\rm I}_{N A}  \\
       \Psi^{\rm I}_{N B} \\
       \Psi^{\rm II}_{1 A} \\
       \Psi^{\rm II}_{1 B}  \\
       \cdots \\
       \Psi^{\rm II}_{N A}  \\
       \Psi^{\rm II}_{N B} 
     \end{bmatrix}
     =
     E\Psi.
  \end{equation}
By solving Eq. (\ref{A-1}), the bulk equation of the chain I can be obtained as follows:
\begin{equation}\label{A-2}
  \begin{aligned}
  E \Psi^{\rm I}_{n,A} = t_2 \Psi^{\rm I}_{n-1,B} + {\rm i}\epsilon \Psi^{\rm I}_{n,A} + (t_1+\gamma) \Psi^{\rm I}_{n,B}, \quad
  E \Psi^{\rm I}_{n,B} = t_2 \Psi^{\rm I}_{n+1,A} + {\rm i}\epsilon \Psi^{\rm I}_{n,B} + (t_1-\gamma) \Psi^{\rm I}_{n,A} .
  \end{aligned}
\end{equation}

For computational convenience, we set the energy  as $E^{\rm I} = E-{\rm i}\epsilon$. This substitution allows Eq. (\ref{A-2}) to be simplified through the replacement of $E$  with $E^{\rm I}$:
\begin{equation}\label{A-3}
  \begin{aligned}
    E^{\rm I} \Psi^{\rm I}_{n,A} = t_2 \Psi^{\rm I}_{n-1,B} + \Psi^{\rm I}_{n,A} + (t_1+\gamma) \Psi^{\rm I}_{n,B}, \quad
    E^{\rm I} \Psi^{\rm I}_{n,B} = t_2 \Psi^{\rm I}_{n+1,A} + \Psi^{\rm I}_{n,B} + (t_1-\gamma) \Psi^{\rm I}_{n,A} .
  \end{aligned}
\end{equation}

The bulk equation for chain II can be obtained in a similar way as follows:
 \begin{equation}\label{A-4}
  \begin{aligned}
    E^{\rm II} \Psi^{\rm II}_{n,A} = t_2 \Psi^{\rm II}_{n-1,B} + \Psi^{\rm II}_{n,A} + (t_1+\gamma) \Psi^{\rm II}_{n,B}, \quad
    E^{\rm II} \Psi^{\rm II}_{n,B} = t_2 \Psi^{\rm II}_{n+1,A} + \Psi^{\rm II}_{n,B} + (t_1-\gamma) \Psi^{\rm II}_{n,A},
  \end{aligned}
\end{equation} 
where $E^{\rm II} = E+{\rm i}\epsilon$.
 
Since the bulk states are exponentially localized at the two domain walls, we can ansatz the eigenstates on chains I and II  to be $ \ket{\Psi^{\rm I}}=\sum_{i} \ket{\phi^{(i)}} $ and $ \ket{\Psi^{\rm II}}=\sum_{i} \ket{\varphi^{(i)}} $, respectively,  where $i$  is the index of the components of the independent wave function.  $\ket{\phi^{(i)}}$ and $\ket{\varphi^{(i)}}$ possess exponential form, and can be expressed as $(\phi_{n,A},\phi_{n,B})= (\beta^{\rm I})^{n}(\phi_{A},\phi_{B})$ and $(\varphi_{n,A},\varphi_{n,B})= (\beta^{\rm II})^{n}(\varphi_{A},\varphi_{B})$ with $\beta=e^{ik+i\kappa}$ the non-Bloch factor, indicating that the momentum is pluralized.

Substituting the ansatz wave function into the bulk quation, there are:
\begin{equation}\label{A-5}
  \begin{aligned}
   (t_1 + \gamma)[(\beta^{\rm I})^n \phi_B] + t_2 (\beta^{\rm I})^{n-1} \phi_B = E^{\rm I} (\beta^{\rm I})^n \phi_A, \quad 
   (t_1 - \gamma)[(\beta^{\rm I})^n \phi_A] + t_2 (\beta^{\rm I})^{n+1} \phi_A = E^{\rm I} (\beta^{\rm I})^n \phi_B,
  \end{aligned}
\end{equation}
\begin{equation}\label{A-6}
  \begin{aligned}
   (t_1 + \gamma)[(\beta^{\rm II})^n \varphi_B] + t_2 (\beta^{\rm II})^{n-1} \varphi_B = E^{\rm II} (\beta^{\rm II})^n \varphi_A, \quad 
   (t_1 - \gamma)[(\beta^{\rm II})^n \varphi_A] + t_2 (\beta^{\rm II})^{n+1} \varphi_A = E^{\rm II} (\beta^{\rm II})^n \varphi_B.
  \end{aligned}
\end{equation}

Eq. (\ref{A-5}) and Eq. (\ref{A-6}) are solvable only if the coefficient determinant is 0, therefore, characteristic equation (as is shown Eq. (\ref{A-7}) and Eq. (\ref{A-8}) ) of chain I and II can be obtained by solving the coefficient determinant of Eq. (\ref{A-5}) and Eq. (\ref{A-6}):
\begin{equation} \label{A-7}
(E^{\rm I})^2=t_1^2-\gamma^2+t_2^2+t_2(\beta^{\rm I})^{-1}(t_1-\gamma)+t_2 \beta^{\rm I} (t_1+\gamma),
\end{equation}
\begin{equation} \label{A-8}
(E^{\rm II})^2=t_1^2-\gamma^2+t_2^2+t_2(\beta^{\rm II})^{-1}(t_1-\gamma)+t_2 \beta^{\rm II} (t_1+\gamma),
\end{equation}
which are quadratic equation  with respect to $\beta^{\rm I}$ and $\beta^{\rm II}$ with two roots, meaning that the wave function index $i=1,2$, and the wave function components of sublattice are satisfied:
\begin{equation}\label{A-9}
 \begin{aligned}
    \phi_A^{(i)}= \frac{(t_1+\gamma)+t_2(\beta^{\rm I}_{i})^{-1}}{E^{\rm I}}\phi_B^{(i)}= \frac{E^{\rm I}}{(t_1-\gamma)+t_2(\beta^{\rm I}_{i})}\phi_B^{(i)} , \quad
    \phi_B^{(i)}=\frac{E^{\rm I}}{(t_1+\gamma)+t_2(\beta^{\rm I}_{i})^{-1}}\phi_A^{(i)}= \frac{(t_1-\gamma)+t_2(\beta^{\rm I}_{i})}{E^{\rm I}}\phi_A^{(i)} .
  \end{aligned}
\end{equation}
\begin{equation}\label{A-10}
\begin{aligned}
    \varphi_A^{(i)}= \frac{(t_1+\gamma)+t_2(\beta^{\rm II}_{i})^{-1}}{E^{\rm II}}\varphi_B^{(i)}= \frac{E^{\rm II}}{(t_1-\gamma)+t_2(\beta^{\rm II}_{i})}\varphi_B^{(i)} , \quad
    \varphi_B^{(i)}=\frac{E^{\rm II}}{(t_1+\gamma)+t_2(\beta^{\rm II}_{i})^{-1}}\varphi_A^{(i)}= \frac{(t_1-\gamma)+t_2(\beta^{\rm II}_{i})}{E^{\rm II}}\varphi_A^{(i)} .
  \end{aligned}
\end{equation}
Eq. (\ref{A-7}) and Eq. (\ref{A-8}) possess two roots means that there are two independently propagating exponential wave functions will be linearly superimposed in a certain form to satisfy the corresponding boundary conditions. Thus, the wave function forms of chains I and II can be expressed in real space as:
\begin{equation}\label{A-11}
  \begin{pmatrix}
    \Psi^{\rm I}_{n,A} \\
    \Psi^{\rm I}_{n,B} 
  \end{pmatrix}
  = (\beta_1^{\rm I})^n
  \begin{pmatrix}
    \phi^{(1)}_A \\
    \phi^{(1)}_B 
  \end{pmatrix}
   +
   (\beta_2^{\rm I})^n
  \begin{pmatrix}
    \phi^{(2)}_A \\
    \phi^{(2)}_B 
  \end{pmatrix},
\end{equation}
\begin{equation}\label{A-12}
  \begin{pmatrix}
    \Psi^{\rm II}_{n,A} \\
    \Psi^{\rm II}_{n,B} 
  \end{pmatrix}
  = (\beta_1^{\rm II})^n
  \begin{pmatrix}
    \varphi^{(1)}_A \\
    \varphi^{(1)}_B 
  \end{pmatrix}
   +
   (\beta_2^{\rm II})^n
  \begin{pmatrix}
    \varphi^{(2)}_A \\
    \varphi^{(2)}_B 
  \end{pmatrix}.
\end{equation}
Here, we set 
\begin{equation}\label{A-13}
\begin{aligned}
 \frac{\phi_A^{(i)}}{\phi_B^{(i)}} &=\frac{(t_1+\gamma)+t_2(\beta^{\rm I}_{i})^{-1}}{E^{\rm I}}= \frac{E^{\rm I}}{(t_1-\gamma)+t_2(\beta^{\rm I}_{i})} = \eta_i^{\rm I}, \quad
 \frac{\varphi_A^{(i)}}{\varphi_B^{(i)}}&=\frac{(t_1+\gamma)+t_2(\beta^{\rm II}_{i})^{-1}}{E^{\rm II}}=\frac{E^{\rm II}}{(t_1+\gamma)+t_2(\beta^{\rm II}_{i})^{-1}}
  = \eta_i^{\rm II}.
\end{aligned}
\end{equation}

\section{Derivation of The Edge Equation}\label{app:B}
The edge conditions (Eq. (\ref{B-1})) can be obtained by solving Eq. (\ref{A-1})
\begin{equation}\label{B-1}
  \begin{aligned}
     &(t_1+\gamma)\Psi^{\rm I}_{1,B} + t_2\Psi^{\rm II}_{N,B}=E^{\rm I}\Psi^{\rm I}_{1,A},  
     &\quad (t_1-\gamma)\Psi^{\rm I}_{N,A} + t_2\Psi^{\rm II}_{1,A}=E^{\rm I}\Psi^{\rm I}_{N,B},\\
     &(t_1+\gamma)\Psi^{\rm II}_{1,B} + t_2\Psi^{\rm I}_{N,B}=E^{\rm II}\Psi^{\rm II}_{1,A},  
    & \quad  (t_1-\gamma)\Psi^{\rm II}_{N,A} + t_2\Psi^{\rm I}_{1,A}=E^{\rm II}\Psi^{\rm II}_{N,B}.\\
  \end{aligned}
\end{equation}
Substituting the wave function in real space (Eq. (\ref{A-9}) and (\ref{A-10}))  into the edge quation, there are:  
\begin{equation}\label{B-2}
  \begin{aligned}
  &(t_1+\gamma)[\beta^{\rm I}_{1} \phi_B^{(1)} + \beta^{\rm I}_{2} \phi_B^{(2)}] + t_2[(\beta^{\rm II}_{1})^N \varphi_B^{(1)} + (\beta^{\rm II}_{2})^N \varphi_B^{(2)}] 
  =  E^{\rm I}[\beta^{\rm I}_{1} \phi_A^{(1)} + \beta^{\rm I}_{2} \phi_A^{(2)}],\\
  &(t_1-\gamma)[(\beta^{\rm I}_{1})^N \phi_A^{(1)} + (\beta^{\rm I}_{2})^N \phi_A^{(2)}] + t_2[\beta^{\rm II}_{1} \varphi_A^{(1)} + \beta^{\rm II}_{2} \varphi_A^{(2)}] 
  =  E^{\rm I}[(\beta^{\rm I}_{1})^N \phi_B^{(1)} + (\beta^{\rm I}_{2})^n \phi_B^{(2)}],\\
  &(t_1+\gamma)[\beta^{\rm II}_{1} \varphi_B^{(1)} + \beta^{\rm II}_{2} \varphi_B^{(2)}] + t_2[(\beta^{\rm I}_{1})^N \phi_B^{(1)} + (\beta^{\rm I}_{2})^N \phi_B^{(2)}] 
  =  E^{\rm II}[\beta^{\rm II}_{1} \varphi_A^{(1)} + \beta^{\rm II}_{2} \varphi_A^{(2)}],\\
  &(t_1-\gamma)[(\beta^{\rm II}_{1})^N \varphi_A^{(1)} + (\beta^{\rm II}_{2})^N \varphi_A^{(2)}] + t_2[\beta^{\rm I}_{1} \phi_A^{(1)} + \beta^{\rm I}_{2} \phi_A^{(2)}] 
  =  E^{\rm II}[(\beta^{\rm II}_{1})^N \varphi_B^{(1)} + (\beta^{\rm II}_{2})^n \varphi_B^{(2)}].
  \end{aligned}
\end{equation}
Then, in combination with Eq. (\ref{A-13}) , we can obtain:
\begin{equation}\label{B-3}
 \begin{aligned}
   E^{\rm I} \eta^{\rm I}_i =  t_1+\gamma+t_2(\beta^{\rm I}_i)^{(-1)},  \quad 
   E^{\rm II} \eta^{\rm II}_i =  t_1+\gamma+t_2(\beta^{\rm II}_i)^{(-1)}.
  \end{aligned}
\end{equation} 
Substituting Eq. (\ref{A-11}), (\ref{A-12}) and (\ref{B-3}) into edge conditions Eq. (\ref{B-2}), then $\varphi_A^{(i)}, \varphi_B^{(i)}$ can be eliminated, getting a set of linear equations be writen $\bm {M\Psi}=0$  as follow:
 \begin{equation}\label{B-4}
   \bm{M}=
     \begin{pmatrix}
      -t_2 & -t_2 & t_2(\beta^{\rm II}_1)^N & t_2(\beta^{\rm II}_2)^N \\
      -t_2 \eta^{\rm I}_1 (\beta^{\rm I}_1)^(N+1) & -t_2 \eta^{\rm I}_2 (\beta^{\rm I}_2)^(N+1) & t_2\eta^{\rm II}_1 \beta^{\rm II}_1 & t_2\eta^{\rm II}_2 \beta^{\rm II}_2 \\
      t_2(\beta^{\rm I}_1)^N & t_2(\beta^{\rm I}_2)^N & -t_2 & -t_2 \\
      t_2 \eta^{\rm I}_1 \beta^{\rm I}_1  & t_2 \eta^{\rm I}_2 \beta^{\rm I}_2 & -t_2 \eta^{\rm II}_1 (\beta^{\rm II}_1)^(N+1)& -t_2 \eta^{\rm II}_2 (\beta^{\rm II}_2)^(N+1) 
    \end{pmatrix},
 \end{equation}
where,$\bm{\Psi}=[\phi_B^{(1)},\phi_B^{(2)},\varphi_B^{(1)},\varphi_B^{(2)}]^{\top}$. The necessary and sufficient condition for the linear equations to have solutions is that the determinant of the coefficient matrix equals zero, i.e., $Det(\bm{M})=0$, it can be calculated as follow:
\begin{equation}\label{B-5}
 \begin{aligned}
 & [(\beta^{I}_1)^N (\beta^{\rm II}_1)^N-1][(\beta^{I}_2)^N (\beta^{\rm II}_2)^N-1](\eta^{I}_1 \beta^{\rm I}_1 -\eta^{\rm II}_2 \beta^{\rm II}_2)(\eta^{\rm I}_2 \beta^{\rm I}_2 -\eta^{\rm II}_1 \beta^{\rm II}_1) \\
  =&[(\beta^{\rm I}_2)^N (\beta^{\rm II}_1)^N-1][(\beta^{\rm I}_1)^N (\beta^{\rm II}_2)^N-1]
  (\eta^{\rm I}_1 \beta^{\rm I}_1 -\eta^{\rm II}_1 \beta^{\rm II}_1)(\eta^{\rm I}_2 \beta^{\rm I}_2 -\eta^{\rm II}_2 \beta^{\rm II}_2) .
 \end{aligned}
\end{equation}

Eq. (\ref{B-5}) is the limiting condition that $\beta^{\rm I / \rm II}$ in the characteristic equation (Eq. (\ref{A-7}) and (\ref{A-8})) need to satisfy under gain and loss domain wall boundary condition.
\section{Derivation of The Continuity Conditions}\label{app:C}

According to non-Bloch theory, the GBZ equation satisfying the continuity condition takes the form $\lvert \beta_m \rvert= \lvert \beta_{m+1} \rvert$, where $\beta$ denotes the Bloch factor and  $m$ indexes the 
$m-{\rm th}$ solution calculated under OBC. Physically, this condition ensures wave function termination at both boundaries, forming a standing wave. However, these conclusions cannot be directly extended to our model for two key reasons: (i) the model resides in an intermediate regime between OBC and PBC, and (ii) the solutions $\beta^{\rm I}$ and $\beta^{\rm II}$ originate from distinct characteristic equations (Eqs. \ref{A-7} and \ref{A-8}), yielding four solutions that resist straightforward ordering. Consequently, the GBZ equation for this model must be reformulated.

We begin by positing that the solutions of the two characteristic equations obey the following fundamental relations:
\begin{equation}\label{C-1}
  \lvert \beta^{\rm I}_2 \rvert \geq \lvert \beta^{\rm I}_1 \rvert, \quad  \lvert \beta^{\rm II}_2 \rvert \geq \lvert \beta^{\rm II}_1 \rvert,
\end{equation}
for simplicity, we set 
\begin{equation}\label{C-2}
\begin{aligned}
  \textsl{g}_1=\lvert \beta^{\rm I}_2\beta^{\rm II}_2 \rvert, \quad \textsl{g}_2=\lvert \beta^{\rm I}_2\beta^{\rm II}_1 \rvert,
 \quad \textsl{g}_3=\lvert \beta^{\rm I}_1\beta^{\rm II}_2 \rvert, \quad \textsl{g}_4=\lvert \beta^{\rm I}_1\beta^{\rm II}_1 \rvert,
\end{aligned}
\end{equation}
consequently, the limiting condition described by Eq. (\ref{B-5}) reduces to:
\begin{equation}\label{C-3}
    [\textsl{g}_4^N-1] [\textsl{g}_1^N-1](\eta^{\rm I}_1 \beta ^{\rm I}_1- \eta^{\rm II}_2 \beta ^{\rm II}_2)(\eta^{\rm I}_2 \beta ^{\rm I}_2- \eta^{\rm II}_1 \beta ^{\rm II}_1)
  = [\textsl{g}_2^N-1][\textsl{g}_3^N-1](\eta^{\rm I}_1 \beta ^{\rm I}_1- \eta^{\rm II}_1 \beta ^{\rm II}_1)(\eta^{\rm I}_2 \beta ^{\rm I}_2- \eta^{\rm II}_2 \beta ^{\rm II}_2).
\end{equation}
When the system approaches the thermodynamic limit, i.e., $ N \to \infty$, the limiting conditions can be simplified as follows:

1. When $\textsl{g}_1 \geq 1, \textsl{g}_2 \geq 1, \textsl{g}_3 \leq 1$ and $\textsl{g}_4 \leq 1$,  $(\textsl{g}_3^N-1) \to -1, (\textsl{g}_4^N-1) \to -1, (\textsl{g}_1^N-1) \to \textsl{g}_1^N, (\textsl{g}_2^N-1) \to \textsl{g}_2^N $, then, divide both sides of Eq. (\ref{C-3}) by $\beta^{\rm I}_2$, then the Eq. (\ref{C-3}) can be simplified to: 
\begin{equation}\label{C-4}
  \begin{aligned} 
   &(\beta ^{\rm II}_2)^N (\eta^{\rm I}_1 \beta ^{\rm I}_1- \eta^{\rm II}_2 \beta ^{\rm II}_2)(\eta^{\rm I}_2 \beta ^{\rm I}_2- \eta^{\rm II}_1 \beta ^{\rm II}_1)
   = (\beta ^{\rm II}_1)^N(\eta^{\rm I}_1 \beta ^{\rm I}_1- \eta^{\rm II}_1 \beta ^{\rm II}_1)(\eta^{\rm I}_2 \beta ^{\rm I}_2- \eta^{\rm II}_2 \beta ^{\rm II}_2),
   \end{aligned}
\end{equation}       
in order to satisfy the continuity conditions, one can be obtained $\lvert \beta ^{\rm II}_2 \rvert = \lvert \beta ^{\rm II}_1 \rvert$  under the thermodynamic limit.

2. When $\textsl{g}_2 \leq 1, \textsl{g}_3 \leq 1 $ and  $\textsl{g}_4 \leq 1$ ,  Eq. (\ref{C-3}) can be simplified to: 
\begin{equation}\label{C-5}
  \begin{aligned}
    \relax{[1-g_1^N]}(\eta^{\rm I}_1 \beta ^{\rm I}_1- \eta^{\rm II}_2 \beta ^{\rm II}_2)(\eta^{\rm I}_2 \beta ^{\rm I}_2- \eta^{\rm II}_1 \beta ^{\rm II}_1)
  = (\eta^{\rm I}_1 \beta ^{\rm I}_1- \eta^{\rm II}_1 \beta ^{\rm II}_1)(\eta^{\rm I}_2 \beta ^{\rm I}_2- \eta^{\rm II}_2 \beta ^{\rm II}_2),
  \end{aligned}
\end{equation}
to satisfy the continuity condition, $\textsl{g}_1=1$ can be obtained, i.e., $\lvert \beta ^{\rm I}_2 \beta ^{\rm II}_2 \rvert -1=0 $, otherwise, if $\textsl{g}_1 \geq 1$ or $\textsl{g}_1 \leq 1$, then  Eq. (\ref{C-3})  becomes:
\begin{equation}\label{C-6}
 (\eta^{\rm I}_1 \beta ^{\rm I}_1- \eta^{\rm II}_2 \beta ^{\rm II}_2)(\eta^{\rm I}_2 \beta ^{\rm I}_2- \eta^{\rm II}_1 \beta ^{\rm II}_1)=0,
\end{equation}
 or 
 \begin{equation}\label{C-7}
  \begin{aligned}
   (\eta^{\rm I}_1 \beta ^{\rm I}_1- \eta^{\rm II}_2 \beta ^{\rm II}_2)(\eta^{\rm I}_2 \beta ^{\rm I}_2- \eta^{\rm II}_1 \beta ^{\rm II}_1)
  = (\eta^{\rm I}_1 \beta ^{\rm I}_1- \eta^{\rm II}_1 \beta ^{\rm II}_1)(\eta^{\rm I}_2 \beta ^{\rm I}_2- \eta^{\rm II}_2 \beta ^{\rm II}_2).
  \end{aligned}
\end{equation}
since $\eta$ is only a function of $\beta^{\rm I}$ and $(\beta^{\rm I})^{-1}$, the above two equations can only solve for a finite number of $\beta$, which is not sufficient to characterize the bulk behavior of the system.

3. When $\textsl{g}_1 \geq 1, \textsl{g}_3 \geq 1 $, $\textsl{g}_2 \leq 1$ and  $\textsl{g}_4 \leq 1$, in the thermodynamic limit, Eq. (\ref{C-3})  becomes: 
\begin{equation}\label{C-8}
  \begin{aligned}
    (\beta ^{\rm I}_2)^N(\eta^{\rm I}_1 \beta ^{\rm I}_1- \eta^{\rm II}_2 \beta ^{\rm II}_2)(\eta^{\rm I}_2 \beta ^{\rm I}_2- \eta^{\rm II}_1 \beta ^{\rm II}_1)
  = (\beta ^{\rm I}_1)^N(\eta^{\rm I}_1 \beta ^{\rm I}_1- \eta^{\rm II}_1 \beta ^{\rm II}_1)(\eta^{\rm I}_2 \beta ^{\rm I}_2- \eta^{\rm II}_2 \beta ^{\rm II}_2),
  \end{aligned}
\end{equation}
and continuity conditions require $\lvert \beta ^{\rm I}_2 \rvert =\lvert \beta ^{\rm I}_1 \rvert$.

4. When $\textsl{g}_1 \geq 1, \textsl{g}_2 \geq 1 $ and  $\textsl{g}_3 \geq 1$, similarly, $\lvert \beta ^{\rm I}_1 \beta ^{\rm II}_1 \rvert -1=0 $ can be obtained.

To sum up, the GBZ  equation of the 1D periodic SSH model chain with gain and loss domain wall is:
\begin{equation}\label{C-9}
0=
  \begin{cases}
    \lvert \beta ^{\rm I}_1 \beta ^{\rm II}_1 \rvert -1 ,  &\textsl{g}_1 \geq 1 \wedge  \textsl{g}_2 \geq 1 \wedge  \textsl{g}_3 \geq 1 \\
    \lvert \beta ^{\rm I}_1 \rvert -\lvert \beta ^{\rm I}_2 \rvert, & \textsl{g}_1 \geq \textsl{g}_3 \geq 1  \geq  \textsl{g}_2 \geq \textsl{g}_4  \\
    \lvert \beta ^{\rm II}_1 \rvert - \lvert \beta ^{\rm II}_2 \rvert, & \textsl{g}_1 \geq \textsl{g}_2 \geq 1 \geq \textsl{g}_3 \geq \textsl{g}_4   \\
    \lvert \beta ^{\rm I}_2 \beta ^{\rm II}_2 \rvert -1,   &\textsl{g}_2 \leq 1 \wedge \textsl{g}_3 \leq 1  \wedge  \textsl{g}_4 \leq 1 \\
  \end{cases}
\end{equation}

Finally, the solutions in the characteristic equation Eq. (\ref{A-7}) and Eq. (\ref{A-8}) that satisfy the above four conditions are the solutions of the GBZ, and the curves formed by them are labeled $C_{\beta}$. By substituting the solution that satisfies the condition into the characteristic equation Eq. (\ref{A-7}) and Eq. (\ref{A-8}), the energy of the system can be obtained, which is taken as the analytical solution. 

 \section{Calculation of The GBZ Equation}\label{app:D}
 According to the characteristic equation (Eq. (\ref{A-7})) of chain I, we set 
\begin{equation}\label{D-1}
  (E^{\rm I})^2=(E-i \epsilon)^2 \equiv f(\beta^{\rm I}),
\end{equation}
where 
\begin{equation}\label{D-2}
  f(\beta^{\rm I})=t_1^2-\gamma^2+t_2^2+t_2(\beta^{\rm I}_i)^{-1}(t_1-\gamma)+t_2 \beta^{\rm I}_i (t_1+\gamma).
\end{equation}
Similarly, for chain II, according to Eq. (\ref{A-8}), there is:
\begin{equation}\label{D-3}
  (E^{\rm II})^2=(E+i\epsilon)^2 \equiv f(\beta^{\rm II}),
\end{equation}
where
\begin{equation}\label{D-4}
  f(\beta^{\rm II})=t_1^2-\gamma^2+t_2^2+t_2(\beta^{\rm II}_i)^{-1}(t_1-\gamma)+t_2 \beta^{\rm II}_i (t_1+\gamma),
\end{equation}
then, the relationship between $E$ and $\beta^{\rm I /\rm II}$ for chain I and II can be written to
\begin{align}
  &E=i \epsilon \pm \sqrt{f(\beta^{\rm I})} \label{D-5}, \\
  &E= -i \epsilon \pm \sqrt{f(\beta^{\rm II})} \label{D-6}.
\end{align}

The solutions satisfying condition 1 ($|\beta^{\rm II}_2|= |\beta^{\rm II}_1|$) can be obtained by setting $\beta^{\rm II}_2= \beta^{\rm II}_1 e^{i\theta}$ with $\theta \in [0,2\pi]$, then substitute it into Eq. (\ref{D-6}), solving the Eq(\ref{D-7}):
\begin{equation}\label{D-7}
  f(\beta^{\rm II}_1) = f(\beta^{\rm II}_1 e^{i\theta}),
\end{equation}
by substituting $\beta^{\rm II}_1$  obtained from Eq. (\ref{D-7}) into Eq. (\ref{D-6}), a series of energies $E$ can be obtained, then, substituting $E$ into Eq. (\ref{D-5}) and (\ref{D-6}) , four sets of $\beta$ can be solved, finally, the solution satisfying condition 1 among them is selected as GBZ.

The solutions satisfying condition 2 ($\beta^{\rm I}_2 \beta^{\rm II}_1 = 0$) can be obtained by setting $\beta^{\rm II}_2= (\beta^{\rm I}_2)^{-1} e^{i\theta}$, then we can solve the following equation:
\begin{equation} \label{D-8} 
   i \epsilon \pm  \sqrt{f(\beta^{\rm I}_2)} = -i \epsilon \pm \sqrt{f((\beta^{\rm I}_2)^{-1}e^{i\theta})},
\end{equation}
in a similar way, by substituting the solution $\beta^{\rm I}_2$  obtained from Eq. (\ref{D-8}) into Eq. (\ref{D-5}), a series of $E$  can be obtained, then, combining with Eq. (\ref{D-5}) and Eq. (\ref{D-6}), four groups of $\beta$ can be solved. Finally, the $\beta$ conforming to condition 1 can be screened as the solution of GBZ. The solutions of conditions 3 and 4 are the same as those of conditions 1 and 2, so they are omitted.

In addition, after the solutions  $\beta$  of GBZ is obtained, the eigenenergy can be obtained by substituting them into the characteristic equation, which is regarded as the analytical solution. Therefore, we can  theoretically predict the energy spectrum of this model.

Next, we mathematically explain why the system evolves from PBC to OBC as the gain and loss domain wall gradually strengthens.  

When $\epsilon\to 0$, Eq. (\ref{D-7}) satisfying condition 1 ($|\beta^{\rm II}_2|= |\beta^{\rm II}_1|$) has no solution, while Eq. (\ref{D-8}) satisfying condition 2 have solutions, and which is simplified to $ f(\beta^{\rm I}_2) =  f((\beta^{\rm I}_2)^{-1}e^{i\theta})$, then, solving this equation and simplifying it, there are:
\begin{equation}\label{D-9}
 t_1(\beta + \beta^{-1} - \beta e^{-i\theta} - \beta^{-1} e^{i\theta}) + \gamma (\beta - \beta^{-1} + \beta e^{-i\theta} - \beta^{-1} e^{i\theta}  ) = 0,
\end{equation}
where $t_1$ and $\gamma$ are arbitrary parameters, for the above equation to hold, then their coefficients must be 0, and therefore, they can be solved $\beta= \pm e^{\frac{i\theta}{2}} $, in this case, the system back BZ.

When $\epsilon$ is large, Eq. (\ref{D-8}) satisfying condition 2 have no solution, while Eq. (\ref{D-7}) satisfying condition 1 ($|\beta^{\rm II}_2|= |\beta^{\rm II}_1|$) have  solutions. We know that $|\beta^{\rm II}_2|= |\beta^{\rm II}_1|$ and $|\beta^{\rm I}_2|= |\beta^{\rm I}_1|$ are the GBZ equations corresponding when chains I  and II are completely disconnected. Then we solve Eq(\ref{D-7}), and the GBZ obtained by solving Eq(\ref{D-7}) in this case is the same as that calculated by the OBC system \cite{PhysRevLett.121.086803}.

In summary, the addition of a strong gain and loss domain wall to a 1D periodic chain is indeed equivalent to a chain originally with PBC being torn into two chains that have the same characteristics as those with OBC.

\renewcommand{\baselinestretch}{0.8}
\twocolumngrid
\bibliography{TuningTopologicalStatesbyDissipation.bib}
\end{document}